\documentclass[trackchanges,preprint2]{aastex701}
\graphicspath{{./}{figures/}}

\begin{document}

\title{VLASS Discovery of a Luminous Galactic Radio Transient Evolving on Decade Timescales}

\author[orcid=0000-0001-8840-922X,sname='Miller']{Jessie M. Miller}
\affiliation{Cahill Center for Astronomy and Astrophysics, California Institute of
Technology, Pasadena, CA 91125, USA.}
\email[show]{jmmiller@astro.caltech.edu}  

\author[orcid=0000-0002-7083-4049]{Gregg Hallinan}
\affiliation{Cahill Center for Astronomy and Astrophysics, California Institute of
Technology, Pasadena, CA 91125, USA.}
\email{gh@astro.caltech.edu}  

\author[orcid=0000-0001-9584-2531]{Dillon Dong}
\affiliation{National Radio Astronomy Observatory, Socorro, NM 87801, USA}
\email{} 

\author[orcid=0000-0002-9540-853X]{Adolfo S. Carvalho}
\affiliation{Center for Astrophysics -- Harvard \& Smithsonian, Cambridge, MA 02138, USA.}
\email{}

\author{S. T. Myers}
\affiliation{National Radio Astronomy Observatory, Socorro, NM 87801, USA}
\email{} 

\author[orcid=0000-0001-8426-5732]{Jean Somalwar}
\affiliation{Department of Astronomy, University of California, Berkeley, CA 94720-3411, USA}
\affiliation{Berkeley Center for Multi-messenger Research on Astrophysical Transients and Outreach (Multi-RAPTOR), University of California, Berkeley, CA 94720-3411, USA}
\affiliation{Kavli Institute for Particle Astrophysics and Cosmology, Stanford, CA 94305, USA}
\email{}  

\author[orcid=0000-0002-4119-9963]{Casey Law}
\affiliation{Cahill Center for Astronomy and Astrophysics, California Institute of Technology, Pasadena, CA 91125, USA.}
\email{} 

\author[orcid=0000-0002-3382-9558]{B. M. Gaensler}
\affiliation{Department of Astronomy and Astrophysics, University of California Santa Cruz, 1156 High Street, Santa Cruz, CA 95064, USA.}
\affiliation{Dunlap Institute for Astronomy and Astrophysics, University of Toronto, 50 St. George Street, Toronto, ON M5S 3H4, Canada.}
\affiliation{David A. Dunlap Department of Astronomy and Astrophysics, University of Toronto, 50 St. George Street, Toronto, ON M5S 3H4, Canada.}
\email{}

\author[]{Vikram Ravi}
\affiliation{Cahill Center for Astronomy and Astrophysics, California Institute of
Technology, Pasadena, CA 91125, USA.}
\email{}

\author[orcid=0000-0002-8400-3705]{Laura Chomiuk}
\affiliation{Center for Data Intensive and Time Domain Astronomy, Department of Physics and
Astronomy, Michigan State University, East Lansing, Michigan 48824, USA.}
\email{}

\author[orcid=0000-0002-5936-1156]{Assaf Horesh}
\affiliation{Racah Institute of Physics, The Hebrew University of Jerusalem, Jerusalem, Israel.}
\email{}  

\author[orcid=0000-0003-3411-6370]{Delina Levine}
\affiliation{Cahill Center for Astronomy and Astrophysics, California Institute of
Technology, Pasadena, CA 91125, USA.}
\email{}  

\author[orcid=0000-0002-8804-3501]{Yuyang Chen}
\affiliation{Center for Astrophysics -- Harvard \& Smithsonian, Cambridge, MA 02138, USA.}
\email{}

\begin{abstract}
We present a multiwavelength analysis of the radio transient VT J1906+0849, discovered as a 70 mJy source in Epoch 1 of the Very Large Array Sky Survey (VLASS), 21 yr after an NRAO VLA Sky Survey (NVSS) non-detection. Radio observations reveal the source was first detected in 2005, peaking at $\gtrsim200$ mJy in 2014, then declining until a late 2025 rebrightening. The transient sits at a Galactic latitude of $\approx0.74^\circ$ and the properties of its optical-infrared counterpart support a Galactic origin. At $d\gtrsim15$ kpc, the extreme radio luminosity is likely powered by sustained accretion onto a compact object. However, a Swift-XRT non-detection shows it is X-ray faint relative to the Galactic X-ray binary population, and the radio emission is distinct from X-ray binaries in its temporal and spectral behavior. Broadband radio spectra suggest synchrotron self-absorption, but size constraints from equipartition and very long baseline interferometry show little to no expansion in the radio-emitting region over 5+ yr, despite significant spectral evolution. Near-infrared spectroscopy reveals a single broad emission line with a stable centroid but variable width and luminosity. We attribute this feature to blueshifted Br$\gamma$ tracing a persistent asymmetric  $\approx2000$ km s$^{-1}$ outflow. These properties are unlike any previously identified Galactic radio source. One possible interpretation is that VT J1906+0849 is a young analog of the microquasar SS 433, with a dense disk wind confining a continuously powered synchrotron outflow. This jet-wind interaction explains the compact, slowly expanding radio source and may contribute to the absence of bright X-ray emission.
\end{abstract}



\section{Introduction}

Transient radio emission is a key tracer of relativistic outflows, shock interactions, and synchrotron-emitting nebulae \citep{1979rpa..book.....R,1982ApJ...259..302C}. Radio wavelengths provide a direct probe of particle acceleration and magnetic field properties, while measurements of spectral shape and temporal evolution constrain source energetics, expansion, and interaction with the surrounding medium (e.g., \citealt{1998ApJ...499..810C,1998ApJ...497L..17S}). In addition, radio emission is largely insensitive to dust extinction, allowing access to otherwise obscured regions of the Galaxy and extragalactic environments.

The discovery space for radio transients has expanded rapidly in recent years with the advent of sensitive, wide-field, multi-epoch surveys. Modern surveys such as the Low Frequency Array Two-metre Sky Survey (LoTSS; \citealt{2017A&A...598A.104S,2019A&A...622A...1S}), the South African Radio Astronomy Observatory MeerKAT International GHz Tiered Extragalactic Exploration Survey (MIGHTEE; \citealt{2016mks..confE...6J}), ThunderKAT (\citealt{2016mks..confE..13F}), and the Australian Square Kilometre Array Pathfinder (ASKAP) Variables and Slow Transients survey (VAST; \citealt{2013PASA...30....6M,2021PASA...38...54M}) have enabled the systematic discovery of slow (timescales $\gtrsim$ seconds) transients. 

A major advance in this landscape is the Karl G. Jansky Very Large Array (VLA) Sky Survey (VLASS; \citealt{2020PASP..132c5001L}), a multi-epoch $2-4$ GHz survey covering $\sim34,000$ deg$^2$ of the northern sky ($\delta\geq-40^\circ$) between 2017 and 2026. VLASS consists of three full-sky epochs (Epochs 1-3) and an additional half-sky epoch (Epoch 4.1) with a cadence of $\sim32$ months, achieving a typical $1\sigma$ sensitivity of $\approx120$ $\mu$Jy beam$^{-1}$ at $2-4$ GHz and an angular resolution of $\lesssim2.5''$ \citep{2020PASP..132c5001L}. The survey will ultimately provide full-Stokes polarization products (Baidoo et al., in prep.), further enhancing its diagnostic capabilities. VLASS has already enabled the discovery of thousands of variable and transient \citep{2025arXiv250800976G} sources, a substantial increase over earlier surveys, which identified only tens of radio transient candidates \citep{2006ApJ...639..331G,2007ApJ...657L..37N,2007PASP..119..122K,2008NewA...13..519K,2009ApJ...704..652N,2009AJ....138..787M,2011ApJ...740...65O,2016ApJ...818..105M}.

The extragalactic transient population has been the focus of most VLASS transient studies thus far. This includes newly launched relativistic jets \citep{2020ApJ...905...74N,2022ApJ...938...43Z}, nuclear transients associated with changes in AGN accretion rates and tidal disruption events \citep[TDEs;][]{2022ApJ...929..184S,2025ApJS..277...50K}, radio-selected candidate TDEs \citep{2022ApJ...925..220R,2023ApJ...945..142S,2025ApJ...985..175S,2025ApJ...982..163S,2025A&A...704A...3K}, a candidate $\gamma$-ray burst orphan afterglow \citep{2018ApJ...866L..22L}, a merger-triggered core-collapse supernova \citep{2021Sci...373.1125D}, an emerging pulsar wind nebula \citep{2023ApJ...948..119D}, and novel searches for jetted extragalactic transients \citep{2025PASP..137h4102S}. 

With the exception of superflares from Solar-type stars detected in VLASS \citep{2026ApJ..1001..143D}, Galactic radio transients remain comparatively underexplored despite strong motivations to identify and characterize these populations. The radio time domain has a long history of serendipitous discoveries in the Milky Way, including pulsars \citep{1968Natur.217..709H}, apparent superluminal motion in relativistic jets \citep{1994Natur.371...46M}, and, more recently, long period radio transients \citep{2022Natur.601..526H}. These discoveries did more than expand the census of radio sources: they revealed compact object physics in otherwise inaccessible regimes, including coherent magnetospheric emission from neutron stars \citep{1969ApJ...157..869G,1975ApJ...196...51R}, relativistic beaming and jet launching in accreting binaries \citep{1999ARA&A..37..409M,2004MNRAS.355.1105F}, and coherent radio emission operating on unexpectedly long timescales \citep{2023Natur.619..487H,2024MNRAS.533.2133C}.

\begin{figure*} 
    \centering
    \includegraphics[width=\linewidth]{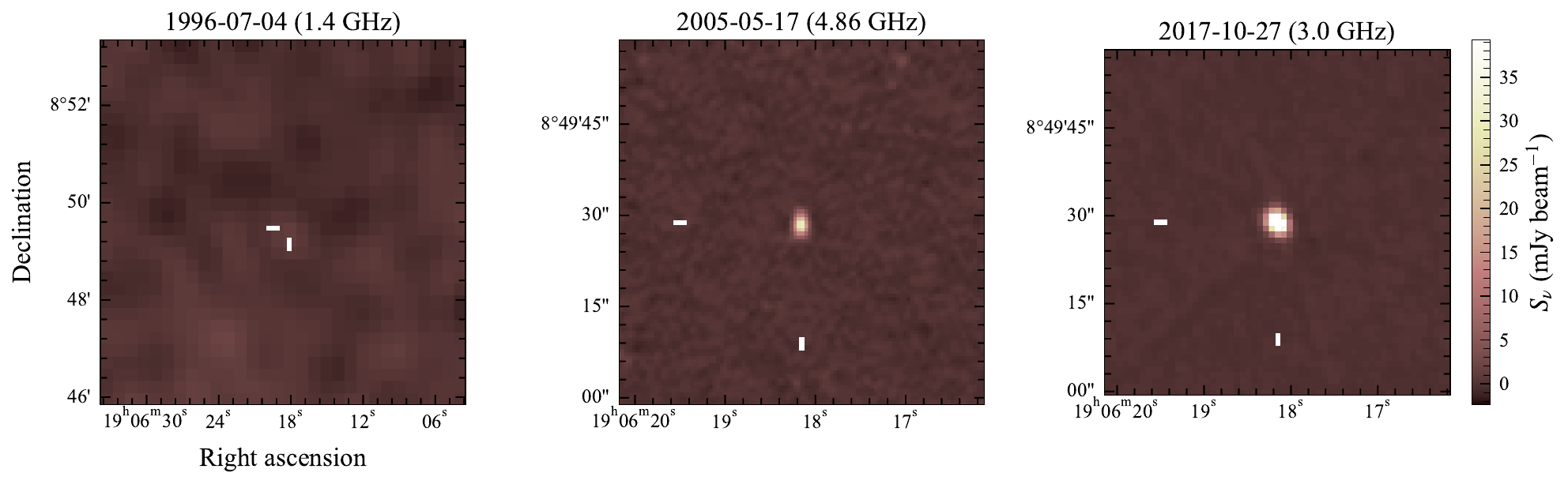}
    \caption{\textit{Left to right}: VT J1906+0849 during the reference epoch (NVSS; $4'\times4'$ field of view), first archival detection (MAGPIS-6cm; $1''\times1''$), and discovery epoch (VLASS Epoch 1; $1''\times1''$). All images share a common flux scale. Attempts to fit a 2D Gaussian at the transient position in NVSS using CASA do not converge.}
    \label{fig:discovery}
\end{figure*}

In parallel, several well-established Galactic populations produce transient radio emission through synchrotron processes, including X-ray binaries in the low/hard state \citep{2001MNRAS.322...31F,2003MNRAS.345.1057M,2006csxs.book..381F}, novae \citep{1986ApJ...305L..71H,2006Natur.442..279O,2014Natur.514..339C}, magnetic cataclysmic variables \citep{1988ApJ...324..431B}, and magnetically active stars \citep{2005A&A...436..241S} and binaries \citep{1987MNRAS.229..659S}. These systems exhibit variability on timescales of days to years and span a wide range of luminosities \citep{2015MNRAS.446.3687P}, making them key targets for systematic identification in modern multi-epoch radio surveys.

However, systematic searches for Galactic radio transients, and the subsequent classification of newly identified sources, are inherently challenging. In many cases, poorly constrained distances make it difficult to infer luminosities, energetics, and physical scales. Multiwavelength follow-up can provide essential classification information by identifying counterparts and constraining the nature of the emitting system, but this effort is complicated in the Galactic plane by severe and spatially variable dust extinction \citep{1989ApJ...345..245C,1998ApJ...500..525S,2011ApJ...737..103S,2019ApJ...887...93G}.

Here we present the discovery and multiwavelength characterization of VLASS Transient J190618.15+084928.85 (hereafter VT J1906+0849), also independently identified by \cite{2025ApJ...987..170C}. In \S\ref{sec:observations}, we identify a multiwavelength counterpart to the radio transient and describe the archival and follow-up observations used to characterize the source. In \S\ref{sec:analysis}, we constrain the distance to the source and the dominant radio emission mechanism, estimate the energetics required to produce the transient, characterize the optical and infrared counterpart, and disfavor an extragalactic origin. In \S\ref{sec:discussion}, we discuss the physical origin of the radio transient and its multiwavelength counterpart. 

\subsection{Discovery in VLASS}
In a search for radio transients at low Galactic latitude in VLASS, we discovered VT J1906+0849, the brightest known transient in VLASS. The source is identified as a transient by detection in VLASS Epoch 1 ($S_\nu\sim70$ mJy at 3 GHz on 2017 October 27) and non-detection in the NRAO VLA Sky Survey (NVSS; \citealt{1998AJ....115.1693C}) on 1996 July 4.  The $3\sigma$ upper limit at the location of VT J1906+0849 in NVSS is $\leq1.8$ mJy at 1.4 GHz. Extrapolating to 3 GHz under the assumption of an optically thick source ($S_\nu\propto\nu^{2.5}$) yields an upper limit of 12.0 mJy. We therefore infer a variability amplitude of $\gtrsim6$ over the $\sim21$ yr interval between the reference and discovery epochs (Figure \ref{fig:discovery}).

The best constraint on source position, $19^{\mathrm h}06^{\mathrm m}18.154227^{\mathrm s}$, $+08^\circ49'28\farcs8560$ (J2000) with 0.1 and 0.6 mas uncertainty in right ascension and declination, is estimated from re-analysis of a Very Long Baseline Array (VLBA) 8.4 GHz observation acquired on 2010 April 27 (\S\ref{sec:br145ao_reduction}).

\section{Observations}\label{sec:observations}

\subsection{Radio}
\subsubsection{Archival interferometric detections}

We inspected archival radio data at the transient position and identified 30+ detections in wide-field radio surveys since 2005. These include: 5 VLA detections (Multi-Array Galactic Plane Imaging Survey ``MAGPIS" 6- and 20-cm, \citealt{2006AJ....131.2525H}; Co-Ordinated Radio `N' Infrared Survey for High-mass star formation ``CORNISH", \citealt{2013yCat..22050001P}; the HI/OH/Recombination line survey of the inner Milky Way ``THOR", \citealt{2018AA...619A.124W}; GLObal view on STAR formation in the Milky Way ``GLOSTAR", \citealt{2024AA...689A.196M}); 1 MeerKAT detection \citep[SARAO MeerKAT Galactic Plane Survey ``SMGPS",][]{2024MNRAS.531..649G}; 31 ASKAP detections (Rapid ASKAP Continuum Survey ``RACS'' low frequency, \citealt{2021PASA...38...58H}; RACS-mid, \citealt{2024PASA...41....3D}; RACS-high, \citealt{2025PASA...42...38D}; VAST, \citealt{2013PASA...30....6M}); and 1 LoTSS detection \citep{Shimwell2026LoTSSDR3}. 

Only a handful of archival VT J1906+0849 detections were supported by published catalog measurements of source flux, position, and fitted size. For all other observations, we use the Common Astronomical Software Applications \citep[CASA;][]{2022PASP..134k4501C} to fit a two-dimensional Gaussian at the transient position. Table \ref{tab:radio_detections} presents all archival radio detections of VT J1906+0849. The transient is unresolved in all detections except in LoTTS, in which it is marginally resolved.

\begin{figure*}
    \centering
    \includegraphics[width=\linewidth]{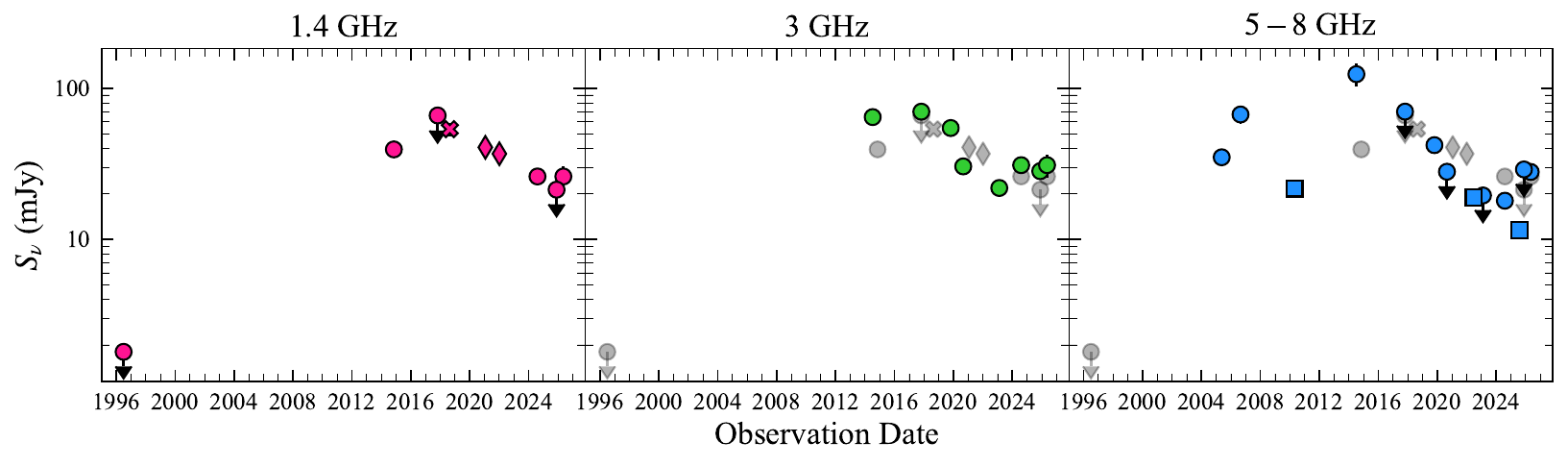}
    \caption{Radio lightcurves of VT J1906+0849. For epochs with in-band (2014, 2017, 2020, 2025) or broadband (2019, 2024, 2026) spectra, we interpolate to 1.4, 3, and 5 GHz where appropriate. Circles, squares, diamonds, and X's correspond to data taken with the VLA, VLBA, ASKAP, and MeerKAT, respectively. The 1.4 GHz lightcurve (\textit{left}) is shown as shaded gray points for comparison in the 3 GHz (\textit{middle}) and $5-8$ GHz (\textit{right}) lightcurves.}
    \label{fig:radio_lightcurves}
\end{figure*}

\begin{figure}
    \centering
    \includegraphics[width=1\linewidth]{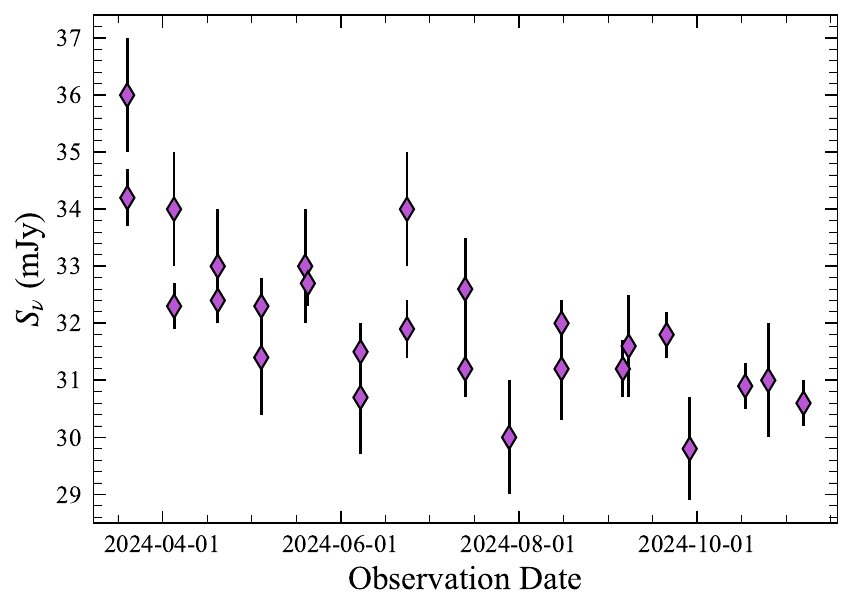}
    \caption{ASKAP VAST 887.5 MHz lightcurve.}
    \label{fig:vast_lightcurve}
\end{figure}

In Figure \ref{fig:radio_lightcurves}, we present the 1.4, 3, and 5-8 GHz lightcurves of the transient. The comparatively densely sampled ASKAP VAST lightcurve is provided in Figure \ref{fig:vast_lightcurve}.



\onecolumngrid
\startlongtable
\begin{deluxetable}{l l c c c c c l} 
\tabletypesize{\footnotesize} 
\tablewidth{0pt}
\tablecaption{Archival radio flux density measurements of compact radio transient VT J1906+0849. Upper limits on $S_\nu$ are reported at the $3\sigma$ level. In the ``Method" column, ``Cat." denotes a published catalog value and ``Ext." indicates source extraction in CASA. 
} 
\label{tab:radio_detections}
\tablehead{
\colhead{Observatory} & \colhead{Survey} & \colhead{Date (Time UTC)} & \colhead{Date (MJD)} & \colhead{$\nu$ (GHz)} &
\colhead{$S_{\nu}$ (mJy)} &
\colhead{Method} & \colhead{Ref.}
}
\startdata
VLA (D) & NVSS & 1996-07-04 & 50633 & 1.4 & $\leq1.8$ &  Ext. & 1 \\  
VLA (D) & VGPS & 2000-08-27 & 51783 &  1.4 & $\leq160$ &  Ext. & 2 \\ 
VLA (B) & MAGPIS-6cm & 2005-05-17 & 53507 &  4.86 & $34.9\pm0.4$ &  Cat. & 3 \\ 
VLA (B) & CORNISH & 2006-08-27 & 53974 &  5 & $67\pm6$ &  Cat. & 4 \\ 
VLBA & BeSSeL calibrators & 2010-04-27 & 55313 &  8.3 & $28$ & Cat. & 5,6 \\
VLA (B,C,D) & MAGPIS-20cm & $<$2013-06-25 & $<$56468 & 1.5 & $29.9\pm0.5$ &  Ext. & 3 \\ 
VLA (D) & GLOSTAR & 2014-07-09 & 56847 &  4.7 & $115\pm6$ &  Cat. & 7 \\ 
VLA (D) & GLOSTAR & 2014-07-09 & 56847 &  5.8 & $150\pm8$ &  Cat. & 7 \\ 
VLA (D) & GLOSTAR & 2014-07-09 & 56847 &  6.9 & $187\pm10$ &  Cat. & 7  \\ 
VLA (C) & THOR & 2014-11-07 & 56968 &  1.06 & $30\pm2$ &  Cat. & 8 \\ 
VLA (C) & THOR & 2014-11-07 & 56968 &  1.13 & $37\pm1$ & Cat. & 8 \\ 
VLA (C) & THOR & 2014-11-07 & 56968 &  1.44 & $39.7\pm0.8$ & Cat. & 8 \\ 
VLA (C) & THOR & 2014-11-07 & 56968 &  1.69 & $47\pm0.9$ & Cat. & 8 \\ 
VLA (C) & THOR & 2014-11-07 & 56968 &  1.82 & $49\pm0.5$ & Cat. & 8 \\ 
VLA (C) & THOR & 2014-11-07 & 56968 &  1.95 & $52\pm0.8$ & Cat. & 8 \\ 
VLA (B) & VLITE & 2017-10-03 & 58029 &  0.340 & $\leq25$ & Ext. & 9, PC \\ 
VLA (B) & VLASS & 2017-10-27 & 58053 &  2.157 & $68.6\pm0.5$ & Ext. & 10 \\ 
VLA (B) & VLASS & 2017-10-27 & 58053 & 2.579 & $70.1\pm0.4$ & Ext. & 10  \\ 
VLA (B) & VLASS & 2017-10-27 & 58053 & 3.048 & $69.9\pm0.4$ & Ext. & 10  \\ 
VLA (B) & VLASS & 2017-10-27 & 58053 & 3.685 & $68.1\pm0.6$ & Ext. & 10  \\
MeerKAT & SMGPS & 2018-08-26 & 58356 & 0.9081 & $53.5\pm0.2$ & Ext. & 11 \\
MeerKAT & SMGPS & 2018-08-26 & 58356 & 0.9524 & $53.5\pm0.2$ & Ext. & 11 \\
MeerKAT & SMGPS & 2018-08-26 & 58356 & 0.9968 & $52.9\pm0.1$ & Ext. & 11 \\
MeerKAT & SMGPS & 2018-08-26 & 58356 & 1.0436 & $53.3\pm0.1$ & Ext. & 11 \\
MeerKAT & SMGPS & 2018-08-26 & 58356 &  1.0929 & $53.4\pm0.1$ & Ext. & 11 \\
MeerKAT & SMGPS & 2018-08-26 & 58356 &  1.1447 & $53.7\pm0.1$ & Ext. & 11 \\
MeerKAT & SMGPS & 2018-08-26 & 58356 &  1.3178 & $53.0\pm0.1$ & Ext. & 11 \\
MeerKAT & SMGPS & 2018-08-26 & 58356 & 1.3817 & $53.4\pm0.1$ & Ext. & 11 \\
MeerKAT & SMGPS & 2018-08-26 & 58356 &  1.4482 & $54.0\pm0.1$ & Ext. & 11 \\
MeerKAT & SMGPS & 2018-08-26 & 58356 &  1.5200 & $55.0\pm0.1$ & Ext. & 11 \\
MeerKAT & SMGPS & 2018-08-26 & 58356 &  1.5944 & $54.5\pm0.1$ & Ext. & 11 \\
MeerKAT & SMGPS & 2018-08-26 & 58356 & 1.6567 & $55.7\pm0.1$ & Ext. & 11 \\
VLA (AnD) & VLITE & 2019-10-26 & 58782 &  0.340 & $19\pm2$ & Ext. & 9, PC \\
ASKAP & RACS-low & 2020-05-03 & 58972 &  0.8875 & $48.2\pm0.8$ & Cat. & 12 \\
VLA (B) & VLITE & 2020-08-31 & 59092 &  0.340 & $\leq100$ & Ext. & 9, PC \\
VLA (B) & VLASS & 2020-08-31 & 59092 &  2.157 & $34.9\pm0.5$ & Ext. & 10  \\
VLA (B) & VLASS & 2020-08-31 & 59092 & 2.579 & $31.9\pm0.3$ & Ext. & 10  \\
VLA (B) & VLASS & 2020-08-31 & 59092 & 3.048 & $30.2\pm0.3$ & Ext. & 10  \\
VLA (B) & VLASS & 2020-08-31 & 59092 &  3.685 & $28.1\pm0.4$ & Ext. & 10  \\
VLA (B) & VLITE & 2020-12-27 & 59210 & 0.340 & $20\pm6$ & Ext. & 9, PC \\
ASKAP & RACS-mid & 2021-01-22 & 59236 & 1.3675 & $40.6\pm0.2$ & Cat. & 13 \\
ASKAP & RACS-high & 2022-01-05 & 59584 &  1.6555 & $36.8\pm0.2$ & Cat. & 14 \\
VLBA & RFC & 2022-06-26 & 59756 & 4.4 &  $23$ & Cat. & 6 \\
VLBA & RFC & 2022-06-26 & 59756 & 7.6 &  $15$ & Cat. & 6 \\
VLA (B) & VLASS & 2023-02-11 & 59986 &  2.157 & $26.1\pm0.4$ & Ext. & 10  \\
VLA (B) & VLASS & 2023-02-11 & 59986 &  2.579 & $23.3\pm0.3$ & Ext. & 10  \\
VLA (B) & VLASS & 2023-02-11 & 59986 &  3.048 & $21.7\pm0.2$ & Ext. & 10  \\
VLA (B) & VLASS & 2023-02-11 & 59986 & 3.685 & $19.5\pm0.4$ & Ext. & 10  \\
VLA (B) & VLITE & 2023-02-11 & 59986 & 0.340 & $\leq22$ & Ext. & 9, PC \\
ASKAP & RACS & 2024-01-04 (04:57) & 60313.20625 & 0.9435 & $32.5\pm0.3$ & Cat. & 16 \\
ASKAP & RACS & 2024-01-04 (05:14) & 60313.21805 & 0.9435 & $33\pm0.8$ & Cat. & 16 \\
ASKAP & VAST & 2024-03-19 (21:45) & 60388.90625 & 0.8875 & $36\pm1$ & Cat. & 15 \\
ASKAP & VAST & 2024-03-19 (23:24) & 60388.975 & 0.8875 & $34.2\pm0.5$ & Cat. & 15 \\
ASKAP & VAST & 2024-04-04 (22:45) & 60404.94791 & 0.8875 & $34\pm1$ & Cat. & 15  \\
ASKAP & VAST & 2024-04-04 (23:25) & 60404.97569 & 0.8875 & $32.3\pm0.4$ & Cat. & 15  \\
ASKAP & VAST & 2024-04-19 (21:44) & 60419.90555 & 0.8875 & $33\pm1$ & Cat. & 15  \\
ASKAP & VAST & 2024-04-19 (22:25) & 60419.93402 & 0.8875 & $32.4\pm0.4$ & Cat. & 15  \\
ASKAP & VAST & 2024-05-04 (19:10) & 60434.79861 & 0.8875 & $31.4\pm1$ & Cat. & 15  \\
ASKAP & VAST & 2024-05-04 (20:05) & 60434.83680 & 0.8875 & $32.3\pm0.5$ & Cat. & 15  \\
ASKAP & VAST & 2024-05-19 (20:02) & 60449.83472 & 0.8875 & $33\pm1$ & Cat. & 15  \\
ASKAP & VAST & 2024-05-20 (17:58) & 60450.74861 & 0.8875 & $32.7\pm0.4$ & Cat. & 15  \\
ASKAP & VAST & 2024-06-07 (18:12) & 60468.75833 & 0.8875 & $30.7\pm1$ & Cat. & 15  \\
ASKAP & VAST & 2024-06-07 (18:54) & 60468.7875 & 0.8875 & $31.5\pm0.5$ & Cat. & 15 \\
ASKAP & VAST & 2024-06-23 (16:04) & 60484.66944 & 0.8875 & $34\pm1$ & Cat. & 15  \\
ASKAP & VAST & 2024-06-23 (16:24) & 60484.68333 & 0.8875 & $31.9\pm0.5$ & Cat. & 15  \\
ASKAP & VAST & 2024-07-13 (14:20) & 60504.59722 & 0.8875 & $32.6\pm0.9$ & Cat. & 15  \\
ASKAP & VAST & 2024-07-13 (14:34) & 60504.60694 & 0.8875 & $31.2\pm0.5$ & Cat. & 15  \\
ASKAP & VAST & 2024-07-28 (15:13) & 60519.63402 & 0.8875 & $30\pm1$ & Cat. & 15  \\
ASKAP & VAST & 2024-08-15 (13:44) & 60537.57222 &  0.8875 & $31.2\pm0.9$ & Cat. & 15  \\
ASKAP & VAST & 2024-08-15 (14:00) & 60537.58333 & 0.8875 & $32\pm0.4$ & Cat. & 15  \\
ASKAP & VAST & 2024-09-05 (13:30) & 60558.5625 & 0.8875 & $31.2\pm0.5$ & Cat. & 15  \\
ASKAP & VAST & 2024-09-07 (12:53) & 60560.53680 &  0.8875 & $31.6\pm0.9$ & Cat. & 15  \\
ASKAP & VAST & 2024-09-20 (12:29) & 60573.52013 &  0.8875 & $31.8\pm0.4$ & Cat. & 15  \\
ASKAP & VAST & 2024-09-28 (10:41) & 60581.44513 &  0.8875 & $29.8\pm0.9$ & Cat. & 15  \\
ASKAP & VAST & 2024-10-17 (10:24) & 60600.43333 &  0.8875 & $30.9\pm0.4$ & Cat. & 15  \\
ASKAP & VAST & 2024-10-25 (7:37) & 60608.31736 &  0.8875 & $31\pm1$ & Cat. & 15  \\
ASKAP & VAST & 2024-11-06 (9:10) & 60620.38194 &  0.8875 & $30.6\pm0.4$ & Cat. & 15  \\
VLA (B) & VLASS & 2025-11-23 & 61002 &  2.157 & $21.3\pm0.6$ & Ext. & 10 \\
VLA (B) & VLASS & 2025-11-23 & 61002 &  2.579 & $26.1\pm0.3$ & Ext. & 10 \\
VLA (B) & VLASS & 2025-11-23 & 61002 &  3.048 & $28.5\pm0.4$ & Ext. & 10 \\
VLA (B) & VLASS & 2025-11-23 & 61002 &  3.685 & $29.1\pm0.4$ & Ext. & 10 \\
LOFAR & LoTSS & 2014-05-23 to 2024-08-19 & 56800-60541 &  0.144 & $11.1\pm0.7$ & Cat. & 17
\enddata
\tablecomments{A machine readable version of this table, including our 19A-422, 24A-404, and 26A-187 VLA follow-up data, is available in the online journal. References: (1) \cite{1998AJ....115.1693C}; (2) \cite{2006AJ....132.1158S}; (3) \cite{2006AJ....131.2525H}; (4) \cite{2013yCat..22050001P}; (5) \cite{2011ApJS..194...25I}; (6) \cite{2025ApJS..276...38P}; (7) \cite{2024AA...689A.196M}; (8) \cite{2018AA...619A.124W}; (9) \cite{2019AAS...23342404P}; (10) \cite{2020PASP..132c5001L}; (11) \cite{2024MNRAS.531..649G}; (12) \cite{2021PASA...38...58H}; (13) \cite{2024PASA...41....3D}; (14) \cite{2025PASA...42...38D}; (15) \cite{2013PASA...30....6M}; (16) \cite{Hotan2020_RACS}; (17) \cite{Shimwell2026LoTSSDR3}. PC indicates private communication.
}
\end{deluxetable}
\twocolumngrid

\subsubsection{Broadband VLA follow-up}
We obtained three broadband VLA follow-up observations of VT J1906+0849. The first observation was carried out on 2019 October 26 (project 19A-422, during A-D reconfiguration) and covered $1-8$ GHz (L-, S-, and C-bands). The second (project 24A-404, B configuration) and third (project 26A-187, A configuration) observations occurred on 2024 August 6 and 2026 May 10, respectively, and spanned 200 MHz to 18 GHz (P-, L-, S-, C-, X-, and Ku-bands\footnote{VLA frequency bands: $200-500$ MHz (P), $1-2$ GHz (L), $2-4$ GHz (S), $4-8$ GHz (C), $8-12$ GHz (X), $12-18$ GHz (Ku).}). In all epochs, 3C 286 was used for flux and bandpass calibration, while J1851+0035 (2019), J1743-0350 (2024), J1922+1530 (2026), and J1851+0035 (2026) served as complex gain calibrators.

Frequency bands above 1 GHz were calibrated using the CASA VLA Calibration Pipeline (v6.6.1 for 19A-422; v6.6.6-18 for 24A-404 and 26A-187), including total electron content (TEC) corrections for $\nu\lesssim4$ GHz. Following pipeline processing, residual radio frequency interference (RFI) was excised via manual flagging and the pipeline was re-run. Phase-only self-calibration was applied once in S- and C-bands in 2019; once in C-band, twice in X- and Ku-bands in 2024; once in each band in 2026. Imaging was performed on a per-spectral-window basis using Briggs weighting \citep{1995PhDT.......238B} with a robustness parameter of 0.5 and Hogbom \citep{1974A&AS...15..417H} deconvolution, with primary beam corrections applied to all images. 

We validate the flux scale in two ways. First, we compare the measured primary beam corrected flux of 3C 286 in every band with its analytic model prediction\footnote{\href{https://science.nrao.edu/facilities/vla/docs/manuals/cal/flux/monitor}{National Radio Astronomy Observatory: Monitoring of Flux Density Calibrators}}. With the exception of Ku-band in 2024 ($\sim20\%$ offset, due to significant flagging), all bands in all epochs are within $\lesssim10\%$ of the model flux. For this reason, we add a 20\% systematic to the 2024 Ku-band detections. Second, we compare two compact, in-field, L-band sources with archival detections and find systematic offsets of $\sim8\%$ (2019) and $\sim10\%$ (2024, 2026), consistent with the $\sim10\%$ systematic uncertainty expected for VLA flux calibration. Program 19A-422 was also accompanied by a $\sim19$ mJy beam$^{-1}$ detection at 340 MHz with the VLA Low-band Ionosphere and Transient Experiment (VLITE; \citealt{2019AAS...23342404P}; private communication).

We manually reduced P-band data, as low frequency calibration is not currently supported by the VLA Calibration Pipeline. After Hanning smoothing, TEC and requantizer gain corrections were applied, followed by automated RFI removal using \texttt{tfcrop}. Delay, bandpass, and gain solutions, and flux scale calibrations, were derived from 3C 286. The calibrated target field was split from the calibration visibility file, and additional RFI removed using \texttt{rflag} and \texttt{extend}. The data were imaged with Briggs weighting ($\rm robust=0$), w-projection (\texttt{wprojplanes}$=-1$), multi-frequency synthesis with two Taylor terms, a single round of phase-only self-calibration, and primary beam correction. 

The P-band flux scale agrees with nearby catalog sources at the few-percent level. VT J1906+0849 is compact in all follow-up VLA images. The corrected peak flux densities for every spectral window are reported in Table \ref{tab:radio_detections}. 

\begin{figure} 
    \centering
    \includegraphics[width=1\linewidth]{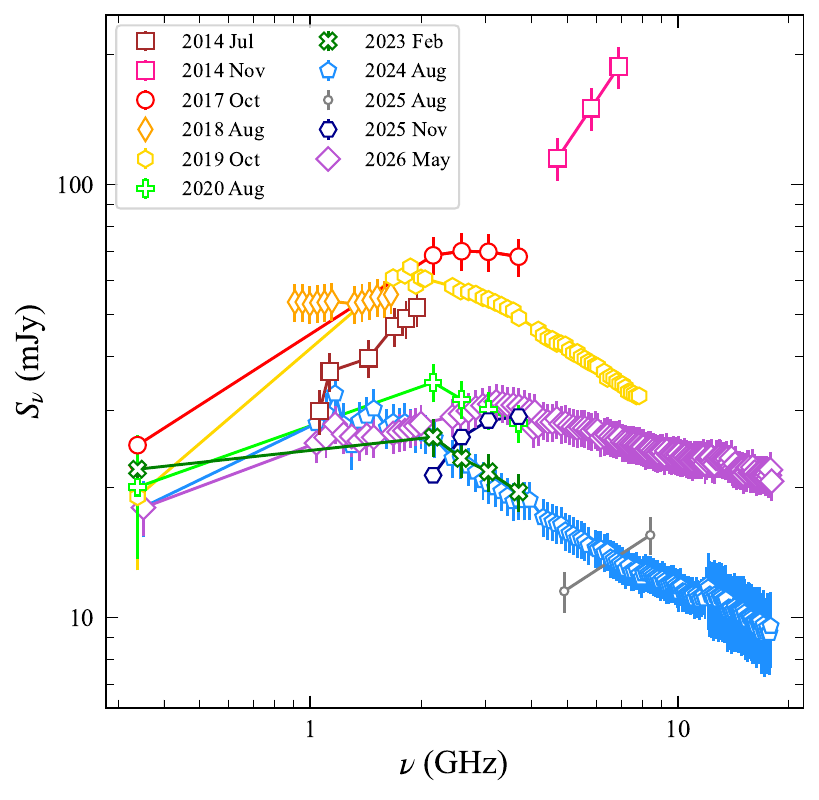}
    \caption{Broadband radio spectra of VT J1906+0849. With the exception of the 2018 August (MeerKAT) and the 2025 August (VLBA), all data were acquired with the VLA.}
    \label{fig:single_epoch_multi_freq_radio}
\end{figure}

Figure \ref{fig:single_epoch_multi_freq_radio} shows the broadband radio spectra of VT J1906+0849 and the flux densities used to construct the 1.4, 3, and $5-8$ GHz lightcurves. We interpolate the spectra to 3 GHz for epochs covering the VLASS band (2014, 2017, 2019, 2020, 2023-2026), to 1.4 GHz for epochs with L-band coverage (2014, 2018, 2024, 2026), and to 5 GHz for epochs extending to $\gtrsim5$ GHz (2014, 2019, 2024, 2026). For VLASS epochs (2017, 2020, 2023, 2025), which lack L- and C-band coverage, we adopt upper limits for the 1.4 and $5-8$ GHz lightcurves by assuming that the flux density outside the VLASS band does not exceed the S-band flux. 

\subsubsection{HI spectral line}
HI 21-cm absorption and emission spectroscopy can constrain the distance to bright continuum emitters (e.g., \citealt{1969ApL.....4..199G}). To this end, we obtained 21-cm spectral line observations toward VT J1906+0489 with the VLA in C configuration on 2025 June 2 and July 29 (project 25A-450). The HI spectral window was centered at 1420.40575 MHz with a bandwidth of 2 MHz divided into 128 channels, corresponding to a channel width of $\sim15.6$ kHz ($\sim3.3$ km s$^{-1}$). The remaining L-band subbands were configured with 64 MHz bandwidths at coarser spectral resolution to improve continuum sensitivity. In both sessions, flux scale and bandpass calibration used 3C 286, while J1856+0610 served as the complex gain calibrator.

Calibration and imaging were performed in CASA. The HI spectral window was extracted from the continuum data, calibrated using standard procedures, and the June and July datasets were combined prior to imaging using \texttt{concat}. Imaging was carried out with \texttt{tclean} in the spectral line cube mode using a 3'' cell size, Briggs weighting ($\rm{robust}=0.3$), and a rest frequency of 1420.40575 MHz. The final data cube used the local standard of rest (kinematic; LSRK) frame and radio velocity convention.

Because the target is compact and the HI line is observed in absorption against the continuum source, large-scale Galactic HI emission must be suppressed to obtain a reliable absorption spectrum for distance constraints. Although C configuration is only sensitive to angular scales up to $\sim16'$, we re-imaged the data using WSClean \citep{offringa-wsclean-2014} with inner \textit{uv}-tapering to filter extended HI emission and isolate the absorption toward the unresolved source. This procedure reduces contamination from diffuse emission and mitigates sidelobe structure near the target position.

\subsubsection{Archival VLBA}\label{sec:br145ao_reduction}


\begin{deluxetable}{rlllllllll}
\tablewidth{0pt}
\tablecaption{VLBA measurements of VT J1906+0849. The upper panel details our reductions of the 2010 April 27 BR145AO and 2025 August 1 BM591a epochs; for BR145AO, we exclude the longest baseline antennas BR, MK, and HN to preserve astrometric accuracy. The lower panel shows 2D Gaussian fits to the published BR145AO (including the longest baselines) and 2022 June 26 BP252 calibrations from \cite{2025ApJS..276...38P}. Columns list observing frequency (GHz), astrometric uncertainty (mas), deconvolved size (mas) and position angle (deg), and integrated (mJy) and peak (mJy beam$^{-1}$) flux densities. Flux density uncertainties are statistical only. 
\label{tab:vlba_fits}}
\tablehead{
\colhead{Year} & \colhead{$\nu$} & \colhead{Position (J2000)}  & \colhead{$\sigma_{\alpha^*},\sigma_{\delta}$} & \colhead{$\theta_{\rm maj}$} & \colhead{$\theta_{\rm min}$} & \colhead{$\theta_{\rm pa}$} & \colhead{$F_{\nu,\rm tot}$} & \colhead{$F_{\nu,\rm pk}$}
}
\startdata
2010 & 8.3 & 19:06:18.154227, +08:49:28.8560 & 0.1, 0.6 & $\leq5.2$ & $\leq2.5$& $--$ & $27\pm3$ & $23\pm2$ \\
2025 & 4.9 & 19:06:18.154179, +08:49:28.855 & 0.1, 2 & $3.2\pm0.2$ & $1.9\pm0.3$ & $120\pm7$ & $25.3\pm0.6$ & $11.5\pm0.2$ \\
2025 & 8.3 & 19:06:18.154209, +08:49:28.855 & 0.1, 1 & $1.4\pm0.2$ & $0.7\pm0.3$ & $120\pm10$ & $24.5\pm0.3$ & $15.5\pm0.1$ \\
\hline
2010 & 8.3 & 19:06:18.1542, +08:49:28.855 & 2 & $1.68\pm0.07$ & $0.32\pm0.2$ & $126\pm3$ & $26.5\pm0.3$ & $21.7\pm0.1$ \\
2022 & 4.4 & 19:06:18.1542, +08:49:28.855 & 2 & $3.7\pm0.2$ & $1.9\pm0.4$ & $7\pm6$ & $23.2\pm0.4$ & $18.9\pm0.2$ \\
2022 & 7.6 & 19:06:18.1542, +08:49:28.855 & 2 & $\leq2.7$ & $\leq1.0$ & $--$ & $15.1\pm0.5$ & $12.2\pm0.2$ \\
\enddata
\end{deluxetable}

The first VLBA detection of VT J1906+0849 was obtained on 2010 April 27 in 8.3 GHz continuum observations as part of the VLBA calibrator search for the BeSSeL survey \citep[project BR145AO;][]{2011ApJS..194...25I}. Several nearby compact sources, including J1905+0952, J1907+0907, and J1906+0838, were also observed, along with the bright fringe finder 1800+782.

Although no detection was reported in the original analysis \citep{2011ApJS..194...25I}, a re-analysis by \cite{2025ApJS..276...38P} reveals a resolved source (Table \ref{tab:vlba_fits}). 

We perform an independent reduction of the BR145AO data to obtain phase-referenced astrometry. Although this was not part of the original observing strategy, we retroactively adopt J1905+0952 as a phase calibrator to define an astrometric reference frame for this epoch. This allows direct comparison with later VLBA observations that use the same calibrator (\S\ref{sec:vlba_followup}). The angular and temporal separations between J1905+0952 and relevant targets are listed in Table \ref{tab:br145ao_positions}.

Data calibration was carried out in CASA (v6.5.4) following standard VLBA procedures, including a priori amplitude calibration, sampler corrections, and fringe fitting using the bright calibrator 1800+782. Eight antennas (BR, FD, HN, KP, MK, NL, OV, and PT) were available during these observations. Due to the $\sim12$ min elapsed between J1905+0952 and VT J1906+0849, significant decorrelation was observed on the longest baselines. We therefore exclude BR, MK, and HN, retaining FD, KP, NL, OV, and PT for calibration and imaging. Phase-only solutions derived from J1905+0952 were applied to the target and check sources.

Imaging was performed with natural weighting and 0.1 mas cell size. The check source J1906+0838 was recovered as a compact point source, while J1907+0907 exhibited residual phase instability consistent with its larger separation from the phase calibrator. VT J1906+0849 was detected and its position measured using \texttt{imfit}, yielding a centroid position of $19^{\mathrm h}06^{\mathrm m}18.154227\pm0.000002^{\mathrm s}$, $+08^\circ49'28\farcs85600\pm0\farcs00006$ (Table \ref{tab:vlba_fits}).

Systematic astrometric uncertainties were estimated by comparing the measured position of the check source J1906+0838 with its reference position \citep{2025ApJS..276...38P}, yielding uncertainties of 0.1 mas in right ascension and 0.6 mas in declination. 

One round of phase-only self-calibration was applied to VT J1906+0849 to mitigate residual phase errors and improve dynamic range. As the source remains unresolved, the synthesized clean beam is adopted as an upper limit on source size. However, we note that the correlated flux density decreases with increasing baseline length (Figure \ref{fig:2025_vlba}), suggesting the source is resolved. The synthesized beam in the image plane is likely insufficient to recover this structure, so Gaussian fitting to the image returns a compact source.


\begin{deluxetable}{rllllllll}
\tablewidth{0pt}
\tablecaption{Sources used in the BR145AO re-analysis. Columns list each source's angular separation from the phase calibrator J1905+0952, $\theta$, and the cadence relative to J1905+0952, expressed as the time separation, $\Delta t$, and number of intervening scans, $N_{\rm s}$.
\label{tab:br145ao_positions}}
\tablehead{
\colhead{Source} & \colhead{Purpose} & \colhead{$\theta$ ($^\circ$)} & \colhead{$\Delta t$ (min)} &  \colhead{$N_{\rm s}$} 
}
\startdata
J1907+0907 & Check source & $0.9$ & $-15.6$ & 4 \\
J1906+0838 & Check source & $1.2$ & $-3.1$ & 0 \\
VT J1906+0849 & Science target & $1.1$ & $+12.7$ & 3 \\
\enddata
\end{deluxetable}

Additional VLBA detections of VT J1906+0849 on 2022 June 26 at 4.4 and 7.6 GHz are also reported by \cite{2025ApJS..276...38P}. While these measurements are not included in our proper motion fitting, we do include their fitted deconvolved Gaussian size and flux density measurements in Table \ref{tab:vlba_fits}.

\subsubsection{VLBA follow-up}\label{sec:vlba_followup}
VT J1906+0849 was observed with the VLBA on 2025 August 1 for a total of three hours using nine antennas (FD, HN, KP, LA, MK, NL, OV, PT, and SC). Observations were conducted with the C- and X-band\footnote{VLBA frequency bands: $3.9-7.9$ GHz (C), $8.0-8.8$ GHz (X).} receivers in the standard dual-frequency setup. The Digital-Down Converter observing system was used in dual polarization mode with $8\times128$-MHz basebands, each subdivided into 256 spectral channels and correlated with the standard DiFX correlator at 2.0-s integration time.

Phase referencing followed a multi-calibrator sequence consisting of J1847+0810 (C1), J1919+0619 (C2), and the primary phase calibrator J1905+0952 (C3), interleaved with scans of the science target (T). All scans followed a C1-C2-C3-T-C1-C2-C3 cadence cycle with a cycle time of approximately 5 minutes. J1924-2914 served as the fringe finder, and J1922+0841 was included as an astrometric check source to assess astrometric systematic uncertainty. 

Data reduction was performed using the Astronomical Image Processing System (AIPS; \citealt{1985daa..conf..195W}) following standard VLBA procedures, with additional steps to implement multi-calibrator phase referencing via the \texttt{ATMCA} task. The calibration sequence included corrections for TEC, Earth orientation parameter, sampler statistics, and parallactic angle. Instrumental delays were solved using a short scan on the bright fringe finder J1924-2914; bandpass calibration and amplitude corrections were performed using the same source. An initial global fringe fit was conducted using J1924-2914 and the primary phase calibrator J1905+0952. The resulting fringe solutions were applied to all sources to generate phase referenced visibilities. 

The secondary phase calibrators were imaged in CASA to assess positional offset relative to the phase center. In X-band, we measure residual position offsets of $+0.341$ mas east and $+0.161$ mas north for J1919+0619, and $-0.398$ mas east and $+0.187$ mas north for J1847+0810. In C-band, we find offsets of $+0.142$ mas east, $+0.390$ mas north for J1919+0619 and $-1.166$ mas east, $-0.096$ mas north for J1847+0810. These offsets are removed using the AIPS task \texttt{CLCOR} prior to solving for scan-based antenna phase solutions on all three phase calibrators with \texttt{CALIB}.

We then applied \texttt{ATMCA}, which models spatial phase gradients using residual phases from all three calibrators and interpolates a tropospheric correction to the target field. The calibrated data were exported to CASA for imaging \citep[\texttt{tclean} with 0.1 mas cell size, natural weighting, and Clark deconvolver;][]{1980A&A....89..377C}. 

\begin{figure*} 
    \centering
    \includegraphics[width=\linewidth]{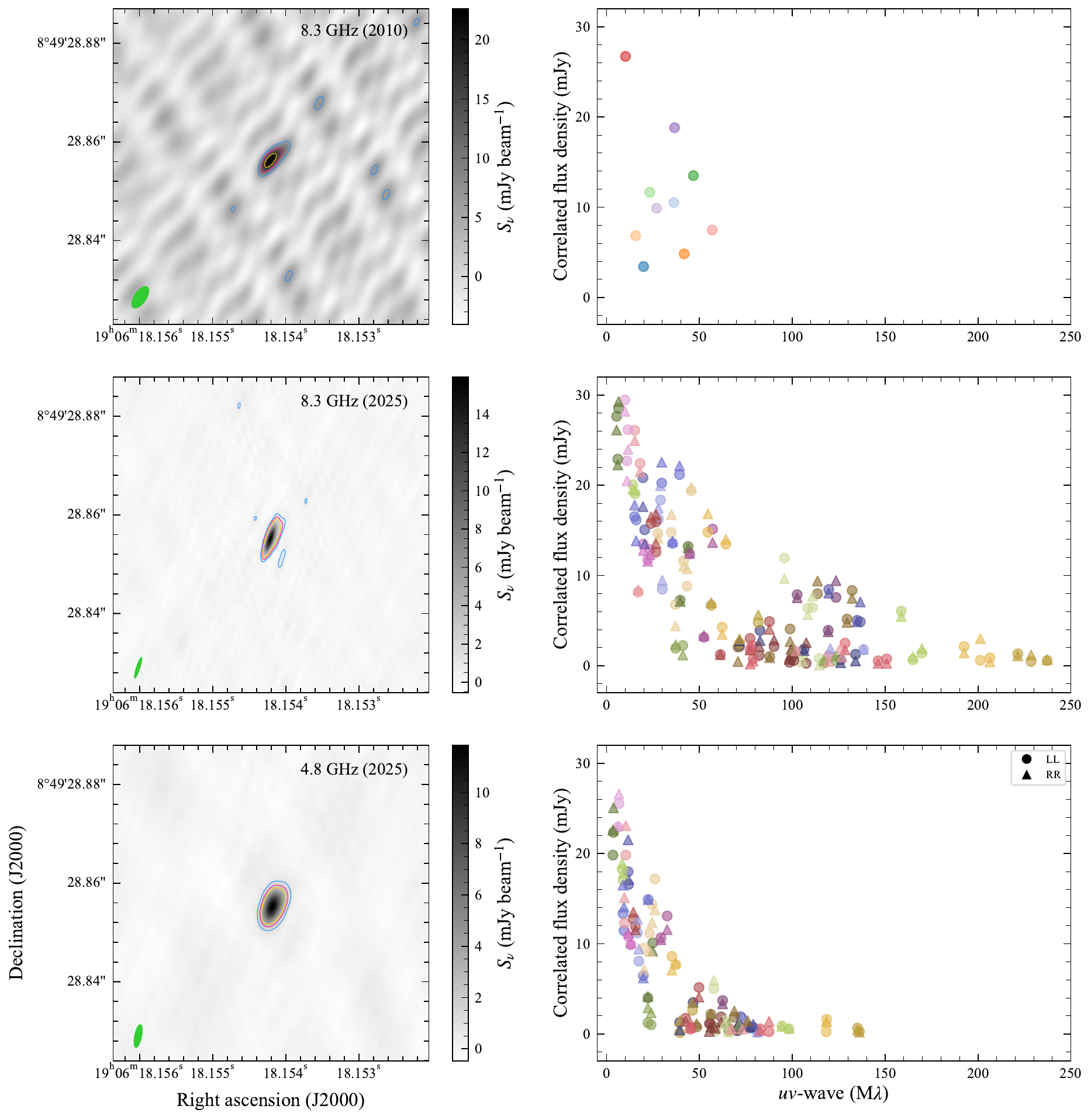}
    \caption{\textit{Top left: }VLBA 8.3 GHz image from 2010 April 27. The restoring beam, $5.2\times2.5$ mas at position angle $\theta_{\rm pa}=-34^\circ$, is shown as a solid green ellipse in the lower-left corner. Contours are drawn at $5\sigma,\ 10\sigma$, and $15\sigma$, where $\sigma=1$ mJy beam$^{-1}$. \textit{Top right: }Correlated flux density as a function of $uv$-wavelength for the left hand (LL) correlations, averaged over all channels and spectral windows, and 3600 s in time. Colors denote unique baselines. \textit{Middle panels: }Same as top panels, but for 2025 August 1 observation at 8.3 GHz, with synthesized beam $4.6\times0.9$ mas, $\theta_{\rm pa}=-18^\circ$, and $\sigma=0.1$ mJy beam$^{-1}$. The $uv$-plot also includes right hand circular polarization (RR). \textit{Bottom panels: }Same as middle panels, but for 2025 August 1 detection at 4.8 GHz, with restoring beam $4.9\times1.6$ mas, $\theta_{\rm pa}=-11^\circ$, and $\sigma=0.2$ mJy beam$^{-1}$.}
    \label{fig:2025_vlba}
\end{figure*}

Following the same procedure as in \S\ref{sec:br145ao_reduction}, we estimate systematic astrometric uncertainties of 0.1 mas in right ascension and 1.0 mas in declination at X-band, and 0.1 mas and 2.0 mas at C-band, respectively. These values reflect the residual calibration uncertainties after phase referencing and are adopted as the dominant source of astrometric error in our analysis. The measured position, total positional uncertainty, deconvolved source size, and total and peak flux densities of VT J1906+0849 are summarized in Table \ref{tab:vlba_fits}. Figure \ref{fig:2025_vlba} illustrates the CLEAN images, and the correlated flux density versus $uv$-wavelength for both C- and X-band detections.

\subsection{Infrared}

\subsubsection{Archival catalog detections}
\begin{figure*}
    \centering
    \includegraphics[width=\linewidth]{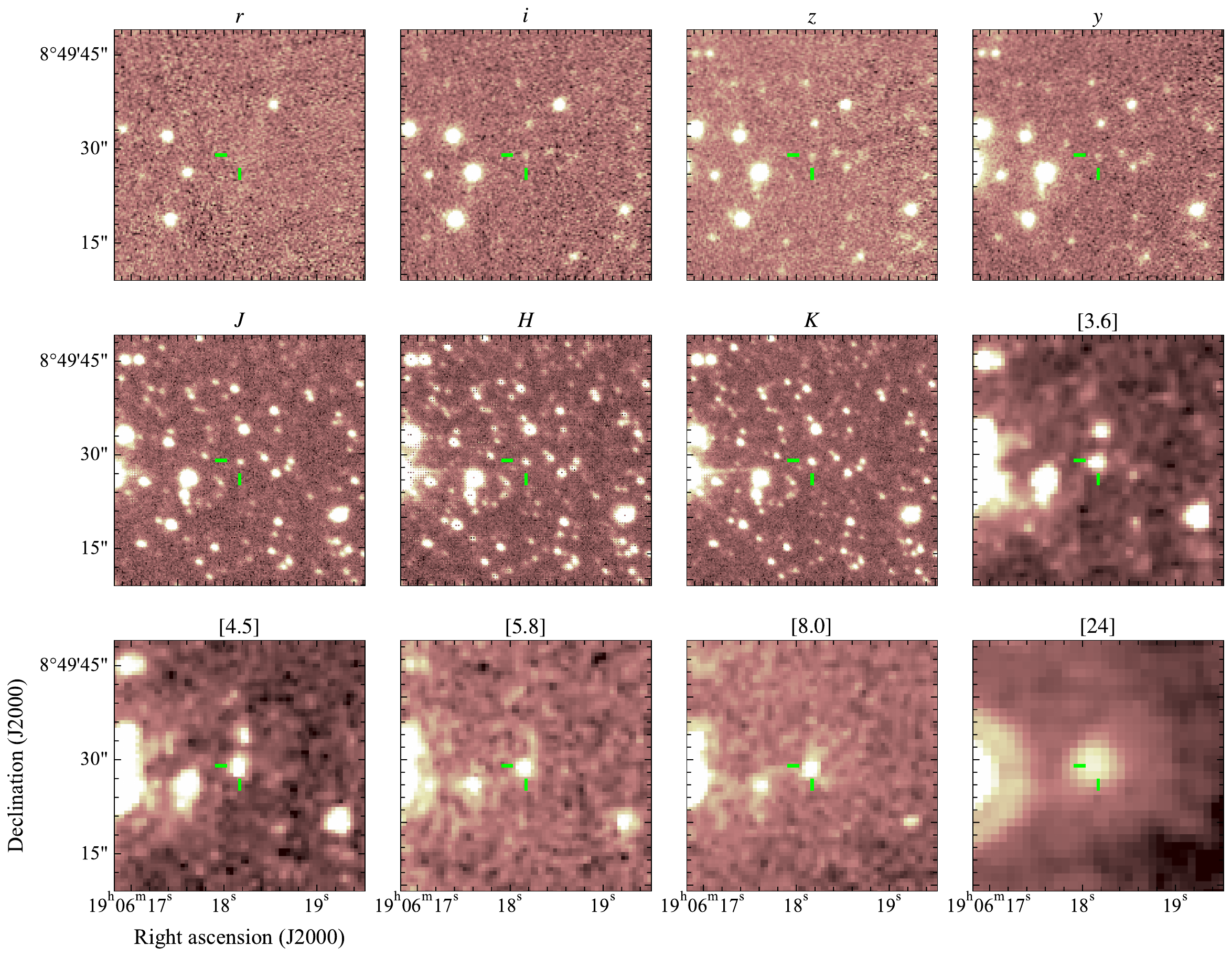}
    \caption{Optical (Pan-STARRS $rizy$), near-infrared (UKIDSS-GPS $JHK$), and mid-infrared (Spitzer/GLIMPSE [3.6], [4.5], [5.8], [8.0]; Spitzer/MIPS [24]) $40''\times40''$ image cutouts centered at the transient position.}
    \label{fig:image_cutouts}
\end{figure*}

Infrared emission spatially coincident with VT J1906+0849 was detected in four surveys: the UKIRT InfraRed Deep Sky Survey-Galactic Plane Survey (UKIDSS-GPS; \citealt{2008MNRAS.391..136L}), the Galactic Legacy Infrared Midplane Survey Extraordinaire (GLIMPSE; \citealt{2009PASP..121..213C}), the unblurred coadds of the Wide-field Infrared Survey Explorer (unWISE; \citealt{2019ApJS..240...30S}), and the Multiband Imaging Photometer for Spitzer (MIPS) Galactic Plane Survey (MIPSGAL; \citealt{2009PASP..121...76C}). The infrared fluxes from these surveys are summarized in Table \ref{tab:photometry}, and Figure \ref{fig:image_cutouts} displays $40''\times40''$ survey image cutouts. 

Assuming a 1 mas total uncertainty in radio position (\S\ref{sec:vlba_followup}), we estimate the probability of chance alignment between the radio and infrared source as $P=N_{\rm IR}\times(\sigma_{\rm tot}/\theta_r)^2$ where $N_{\rm IR}$ is the number of detected infrared sources within a radius $\theta_r=1'$ of the radio source and $\sigma_{\rm tot}=\sqrt{\sigma_{\rm rad}^2+{\sigma_{\rm IR}^2}}$ is the quadrature sum of the uncertainty in radio and infrared positions. We find $10^{-3}$ and $4\times10^{-3}$ in UKIDSS-GPS and GLIMPSE, respectively.


\begin{deluxetable*}{rllllll}
\tablewidth{0pt}
\tablecaption{Photometric detections coincident with VT J1906+0849. 
\label{tab:photometry}}
\tablehead{
\colhead{Date} & \colhead{$\lambda$ ($\mu$m)} & \colhead{Magnitude (Vega)} & \colhead{$S_\nu$ (mJy)} & \colhead{Position (J2000)} & $\sigma$ (mas) & Ref.
}
\startdata
2008-04-15 & 1.25 & $18.46\pm0.06$ & $0.063\pm0.004$ & 19:06:18.17, +08:49:28.81 & 90 & 1 \\  
2008-04-15  & 1.60 & $16.94\pm0.04$ & $0.171\pm0.006$ & 19:06:18.17, +08:49:28.81 & 90 & 1 \\ 
2008-04-15  & 2.20 & $15.65\pm0.02$ & $0.348\pm0.006$ & 19:06:18.17, +08:49:28.81 & 90 & 1 \\ 
2004-10-09 & 3.6 & $13.13\pm0.08$ & $1.5\pm0.1$ & 19:06:18.2, +08:49:28.7 & 400 & 2 \\
2004-10-09 & 4.5 & $12.39\pm0.09$ & $2.0\pm0.2$ & 19:06:18.2, +08:49:28.7 & 400 & 2 \\
2010-10-15 & 4.6 & $12.518\pm0.006$ & $1.69\pm0.01$ & 19:06:18.19, +08:49:29.31 & 30, 50 & 3 \\
2004-10-09 & 5.8 & $11.48\pm0.08$ & $2.9\pm0.2$ & 19:06:18.2, +08:49:28.7 & 400 & 2 \\
2004-10-09 & 8.0 & $10.88\pm0.07$ & $2.8\pm0.2$ & 19:06:18.2, +08:49:28.7 & 400 & 2 \\
2005-10-03 & 24 & $7.81\pm0.07$ & $5.5\pm0.4$ & 19:06:18.2, +08:49:28.9 & 400 & 4 \\ 
\tableline
\multicolumn{5}{c}{} \\[-0.15in]
\tableline
2021-05-29 & 1.25 & $16.9\pm1$ & $0.3\pm0.2$ & - & - & This work\\
2021-05-29 & 1.60 & $16.0\pm0.3$ & $0.4\pm0.1$ & - & - & This work\\ 
2021-05-29 & 2.20 & $15.1\pm0.2$ & $0.56\pm0.08$ & - & - & This work\\ 
\tableline
2024-06-18 & 1.25 & $18.7\pm0.3$ & $0.05\pm0.01$ & - & - & This work\\ 
2024-06-18 & 1.60 & $17.0\pm0.2$ & $0.16\pm0.03$ & - & - & This work\\ 
2024-06-18 & 2.20 & $15.5\pm0.1$ & $0.40\pm0.04$ & - & - & This work\\
\tableline
2025-05-14 & 1.25 & $18.5\pm0.9$ & $0.06\pm0.05$ & - & - & This work\\ 
2025-05-14 & 1.60 & $16.9\pm0.5$ & $0.18\pm0.08$ & - & - & This work\\ 
2025-05-14 & 2.20 & $15.5\pm0.1$ & $0.39\pm0.05$ & - & - & This work\\ 
\tableline
2026-05-08 & 1.25 & $18.0\pm0.3$ & $0.10\pm0.02$ &  - & -  & This work \\
2026-05-08 & 1.60 & $16.6\pm0.2$ & $0.24\pm0.04$ & - & -  & This work \\
2026-05-08 & 2.20 & $15.1\pm0.1$ & $0.59\pm0.07$ &  - & -  & This work \\
\tableline
\multicolumn{5}{c}{} \\[-0.15in]
\tableline
2010-2014 & 0.75 & $22.0\pm0.2$ & $(4.0\pm0.7)\times10^{-3}$ & 19:06:18.15, +08:49:28.97 & 20 & 5, This work \\
2010-2014 & 0.87 & $21.2\pm0.3$ & $(8\pm2)\times10^{-3}$ & 19:06:18.15, +08:49:28.97 & 20 & 5, This work \\
2010-2014 & 0.96 & $20.4\pm0.3$ & $(14.5\pm4)\times10^{-3}$ & 19:06:18.15, +08:49:28.97 & 20 & 5, This work \\
\tableline
2021-5-14 & 0.75 & $21.6\pm0.3$ & $(6\pm2)\times10^{-3}$ & - & - & This work \\
2021-5-14 & 0.87 & $20.3\pm0.2$ & $(18\pm3)\times10^{-3}$ & - & - & This work \\
2021-5-14 & 0.96 & $19.5\pm0.2$ & $(33\pm5)\times10^{-3}$ & - & - & This work \\
\tableline
2023-10-17 & 0.75 & $22.1 \pm 0.3$ & $(3.8\pm0.9)\times10^{-3}$ & - & - & This work\\ 
2023-10-17 & 0.87 & $20.8 \pm 0.2$ & $(11\pm2)\times10^{-3}$ & - & - & This work\\ 
2023-10-17 & 0.96 & $20.1 \pm 0.2$ & $(19\pm3)\times10^{-3}$ & - & - & This work\\ 
\tableline
2025-04-28 & 0.75 & $22.8 \pm 0.3$ & $(2\pm0.6)\times10^{-3}$ & - & - & This work\\ 
2025-04-28 & 0.87 & $21.3 \pm 0.2$ & $(7\pm1)\times10^{-3}$ & - & - & This work\\ 
2025-04-28 & 0.96 & $20.4 \pm 0.2$& $(15\pm3)\times10^{-3}$ & - & - & This work\\ 
\enddata
\tablecomments{Columns list observation date, effective wavelength of the relevant filter, apparent magnitude and uncertainty in the Vega system, flux density and uncertainty, position of the optical/infrared source (if measured using aperture photometry), uncertainty in position $\sigma$, and reference (where applicable). The effective wavelengths of 1.25, 1.60, and 2.20 $\mu$m correspond to the UKIRT/UKIDSS $JHK$ filters; 3.6, 4.5, 5.8, and 8.0 $\mu$m to Spitzer/IRAC bands 1-4; 4.6 $\mu$m to WISE $W2$ filter; 24 $\mu$m to Spitzer/MIPS 24 $\mu$m filter; and 0.75, 0.87, and 0.96 $\mu$m to the Pan-STARRS $izy$ filters. References: (1) \cite{2008MNRAS.391..136L}; (2)  \cite{2009yCat.2293....0S}; (3) \cite{2019ApJS..240...30S}; (4) \cite{2015AJ....149...64G}; (5) \cite{2016arXiv161205560C}.}
\end{deluxetable*}

\subsubsection{Keck-II/NIRES}

We acquired three near-infrared ($0.94-2.45$ micron) spectra using the 18\arcsec\ long, 0\farcs55 slit ($R\sim2700$) on the Near-Infrared Echellete Spectrometer \citep[NIRES,][]{Wilson_NIRES_2004SPIE} on Keck-II. The first observation, on 2021 May 29, consisted of $20\times15$ s exposures (total integration time 300 s) at an airmass of 1.09. The second observation, obtained on 2024 June 18, consisted of $4\times300$ s exposures at an average airmass of 1.03. The third observation, obtained on 2026 May 8, also consisted of $4\times300$s exposures at an airmass of 1.08. 

\begin{figure*}
    \centering
    \includegraphics[width=\linewidth]{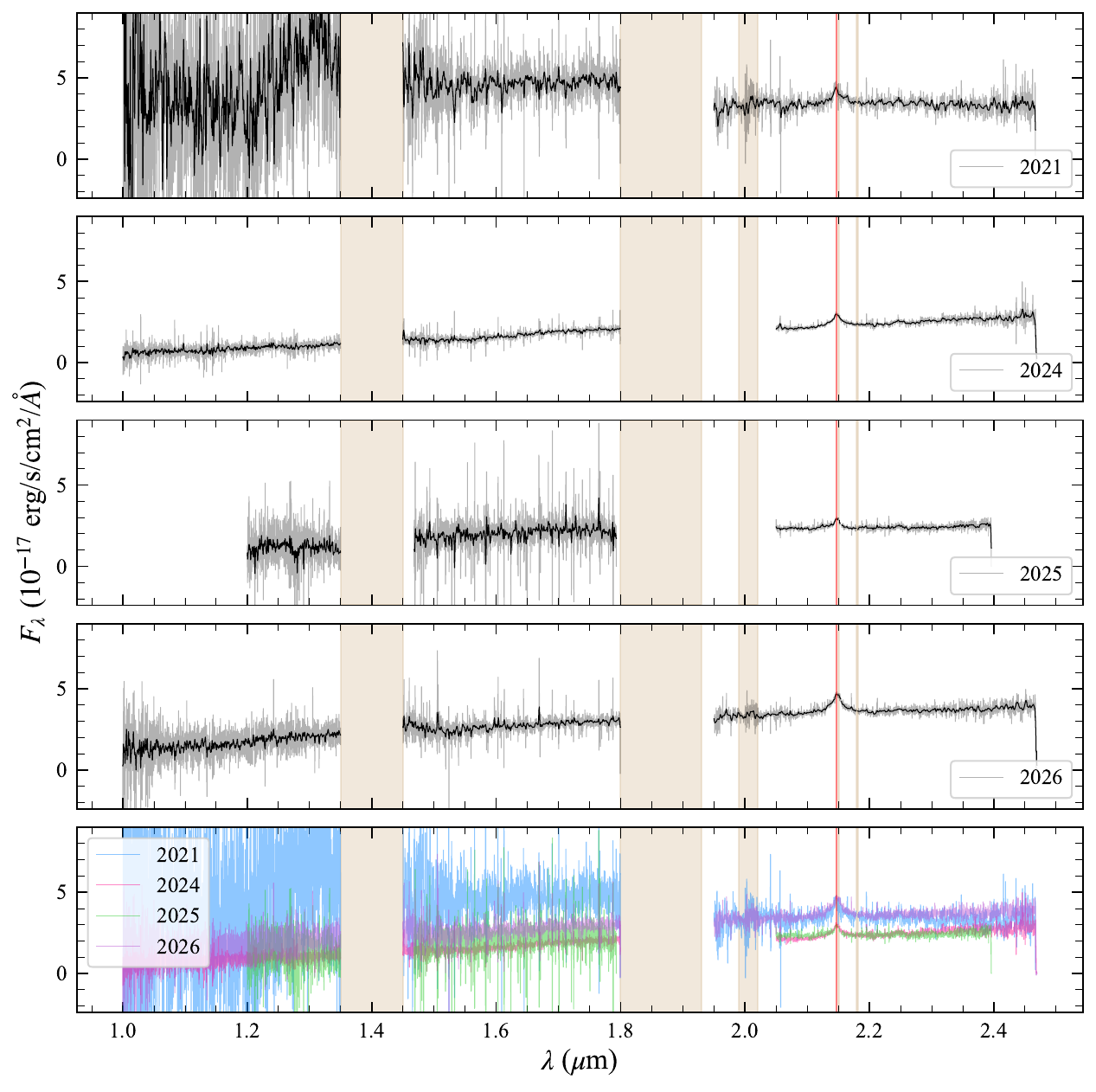}
    \caption{Keck-II/NIRES (2021 May 29, \textit{top}; 2024 June 18, \textit{second panel}; 2026 May 8, \textit{fourth panel}) and Keck-I/MOSFIRE (2025 May 14, \textit{middle}) spectra. All spectra are smoothed using a Gaussian kernel with a standard deviation of 3 \AA. The red solid line (21472 \AA) indicates the broad emission feature and tan shaded regions represent regions of significant telluric absorption. The bottom panel shows all four near-infrared spectra overlaid for comparison.}
    \label{fig:nir_spec}
\end{figure*}

The 2024 and 2026 observations were obtained in an ABBA dither pattern, whereas the 2021 observations were not, resulting in stronger OH residuals in the latter. All three spectra were extracted, co-added, flux calibrated, and telluric corrected using the \texttt{PypeIt} pipeline \citep{2020JOSS....5.2308P,2020zndo...3743493P}. The A0V standard HIP 98640 was used for flux calibration in 2021 and 2024, and HD 121880 was used in 2026. The reduced spectra are shown in Figure \ref{fig:nir_spec}.

\subsubsection{Keck-I/MOSFIRE}
We additionally obtained a near-infrared spectrum using the Multi-Object Spectrograph for Infrared Exploration \citep[MOSFIRE,][]{McLean_MOSFIRE_1_2010SPIE,McLean_MOSFIRE_2_2012SPIE} on Keck-I on 2025 May 14. The observation used the $1''$ longslit ($R\sim3318-3610$) with  exposure times of $8\times120$, $4\times120$, and $4\times180$ s in the $JHK$ filters and an ABBA slit nod dither pattern. The source was observed through significant cloud cover at an airmass of $\sim1$. The MOSFIRE spectra were also fully reduced using \texttt{PypeIt}, with the A0V standard HIP 98640 used for flux calibration.

\subsection{Optical}
\subsubsection{Archival catalog detections}\label{sec:opt_archival}


We inspected archival Pan-STARRS Data Release 1 \citep[PS1;][] {2016arXiv161205560C} stack images at the position of VT J1906+0849. Although the Pan-STARRS catalog reports a source coincident with the radio transient in $r$ and $z$ bands, visual inspection of the stacked images does not support a convincing detection in the $r$ band (Figure \ref{fig:image_cutouts}). A faint source is visible in the $i$, $z$, and $y$ stacks, so we performed independent PSF photometry on these images using the \texttt{photutils} package \citep{larry_bradley_2025_17129028}.

The Pan-STARRS stack images are already sky-subtracted, so we did not apply an additional global background subtraction. For each of the $i$, $z$, and $y$ bands, we identified point sources in the image using \texttt{DAOStarFinder} with a threshold of $6\times$ the sigma-clipped background rms. We selected isolated, unsaturated stars with acceptable sharpness ($0.3-0.9$) and roundness ($<0.5$) values. These cuts were chosen to identify clean stellar sources suitable for estimating the image PSF. The selected stars were used to estimate the stellar full-width at half-maximum (FWHM), which was then used to define a circular Gaussian point response function model for PSF photometry.

We modeled the field sources using iterative PSF photometry. During fitting, a local background was estimated using \texttt{LocalBackground} in an annulus extending from $2\times$ to $4\times$ FWHM from the source position. We retained only well-behaved PSF fits with no \texttt{photutils} flags. We also required $|\texttt{qfit}|\leq2$ and $|\texttt{cfit}|\leq1$, which measure the absolute PSF-fitting and central pixel residuals normalized by the fitted source flux.

At the position of VT J1906+0849, we then performed forced PSF photometry. This target fit used the same circular Gaussian PSF model and included the same local background treatment. The fitted flux was converted to a Pan-STARRS stack magnitude using the exposure time measured from the corresponding auxiliary exposure map over the fitting region\footnote{See ``Photometric Calibration'' section of \href{https://outerspace.stsci.edu/display/PANSTARRS/PS1+Stack+images\#PS1Stackimages-Photometriccalibration}{PS1 Image data products}.}. The statistical magnitude uncertainty was propagated from the \texttt{photutils} flux uncertainty.

To assess the accuracy of our measurements, we cross-matched the well-fit field sources against the Pan-STARRS stack catalog within $1''$. We excluded ambiguous matches with more than one catalog source and sources fainter than the nominal Pan-STARRS $3\pi$ survey limits. For each band, we measured the difference between the catalog PSF magnitude and our PSF-fitting magnitude for the matched field stars. We applied the median offset to the forced photometry of VT J1906+0849, and treated the scatter in the offsets as systematic calibration uncertainty, which was added in quadrature with the statistical fitting uncertainty.

This procedure gives $i=22.4\pm0.2$, $z=21.7\pm0.3$, and $y=21.0\pm0.3$ mag (AB) for VT J1906+0849. These values are also presented in the Vega magnitude system in Table \ref{tab:photometry}.

The positional association between VT J1906+0849 and the PS1 source is unlikely to be due to chance. From the point source extraction described above, we estimate a mean statistical positional uncertainty of $\approx10$ mas for sources in the PS1 images. Combining this term with the 18 mas systematic astrometric uncertainty reported for moderately faint PS1 sources \citep[$i\sim19$;][]{2016arXiv161205560C} gives a total PS1 positional uncertainty of $\approx20$ mas. This dominates the $\approx1$ mas uncertainty of the radio position. We find 173 PS1 sources within 1' of VT J1906+0849, corresponding to a local surface density of 0.015 sources per square arcsec. The probability of finding an unrelated PS1 source within the 20 mas PS1 positional uncertainty is $\sim2\times10^{-5}$.

\subsubsection{Keck-I/LRIS}

\begin{figure*}
    \centering
    \includegraphics[width=\linewidth]{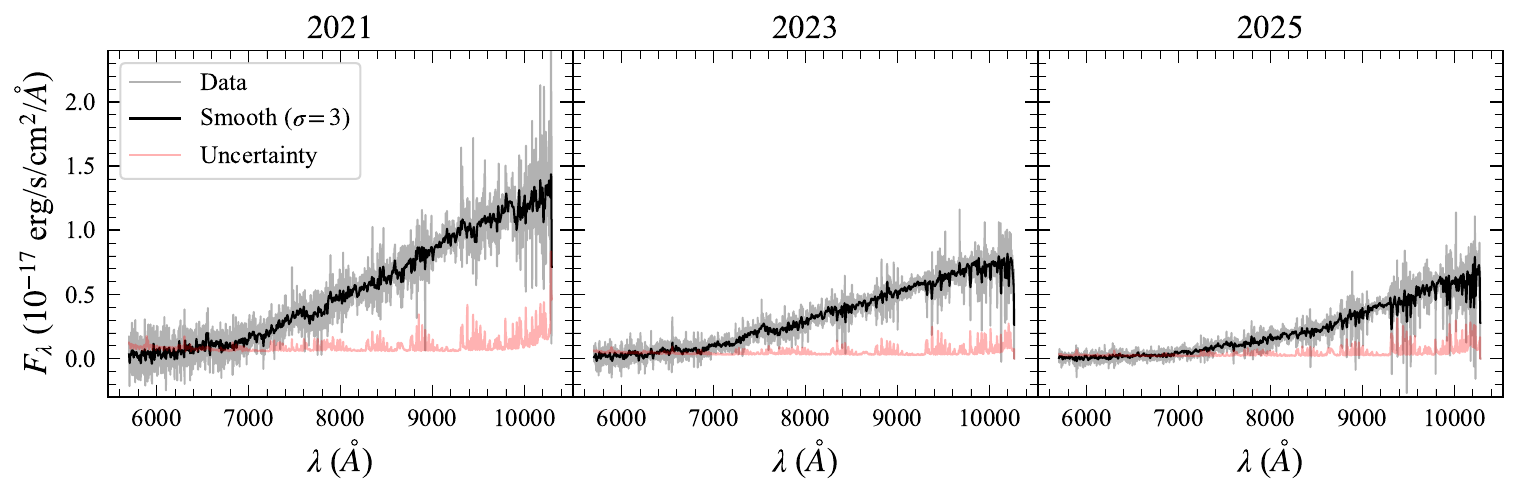}
    \caption{Keck-I/LRIS spectra obtained on 2021 May 14 (\textit{left}), 2023 October 17 (\textit{center}), and 2025 April 28 (\textit{right}). As in Figure \ref{fig:nir_spec}, the spectra are smoothed with a Gaussian kernel and a $\sigma=3$ \AA\ window.}
    \label{fig:opt_spec}
\end{figure*}

We obtained three optical spectra of VT J1906+0849 using the Low Resolution Imaging Spectrograph \citep[LRIS,][]{Oke_LRIS_Red_1995PASP, McCarthy_LRIS_blue_1998SPIE, Rockosi_LRIS_Red_Upgrade_2010SPIE} on Keck-I. The first observation, on 2021 May 14, consisted of $1\times1200$ s on the blue side and $2\times550$ s on the red side at an airmass of 1.03. The second observation, on 2023 October 17, consisted of $4\times900$ s exposures on both the blue and red sides at an average airmass of 1.38. The final observation, on 2025 April 28, consisted of $3\times1200$ s on the blue side and $6\times600$ s on the red side at an average airmass of 1.07.

All observations used the 1\farcs0 slit, clear filter, 560 dichroic, 400/3400 grism ($\rm FWHM\approx6.5-7.1$ \AA) on the blue side, and 400/8500 grating ($\rm FWHM\approx6.9$ \AA) on the red side. The spectra were background subtracted, co-added, extracted, flux calibrated, and telluric-corrected using the \texttt{PypeIt} pipeline and adopting the spectrophotometric standard BD+28d4211 for flux calibration. To ensure robust detection of the trace in each epoch, the 2D spectra were co-added prior to extraction. No trace is detected in any of the blue-side spectra. The 2021, 2023, and 2025 spectra are shown in Figure \ref{fig:opt_spec}.

\subsection{X-ray}
\subsubsection{Swift-XRT}
A 1.5 ks exposure was obtained with the Neil Gehrels Swift Observatory on 2023 June 12, with the Swift X-Ray Telescope (XRT) instrument in photon counting mode (ObsID 00016068001). The data were processed following standard procedures provided in the \cite{2014ascl.soft08004N} v6.32\footnote{\href{https://heasarc.gsfc.nasa.gov/ftools}{HEASoft FTOOLS: A General Package of Software to Manipulate FITS Files}} software suite with the latest Swift-XRT CALDB files. Cleaned event files were retrieved from the UK Swift Science Data Centre\footnote{\href{https://www.swift.ac.uk}{UK Swift Science Data Centre}} and combined using \texttt{XSELECT}. Source and background regions were defined using circular apertures with radii of 20 pixels ($\sim47\arcsec$) and 40 pixels ($\sim94\arcsec$), respectively. The background region was selected from a nearby source-free area on the same CCD.

No X-ray counterpart was detected at the position of VT J1906+0849. Assuming a background-subtracted Poisson distribution, we derive a $3\sigma$ upper limit of $\leq3\times10^{-3}$ ct s$^{-1}$ on the source rate.

\section{Analysis}\label{sec:analysis}
\subsection{Distance estimation}\label{sec:kde}

An HI absorption spectrum toward a compact radio continuum source provides a geometric distance constraint because foreground cold HI clouds absorb the continuum emission, while clouds behind the source contribute only emission. For VT J1906+0849 ($l\approx42.47^\circ,b\approx0.74^\circ$), circular Galactic rotation along this Quadrant I sightline predicts positive local-standard-of-rest velocities, $v_{\rm LSRK}$, for gas inside the Solar Circle ($R<R_0$). Absorption at the largest positive velocity therefore sets a lower limit on the source distance, and absorption approaching the tangent point velocity, $v_{\rm tan}$, implies $d\gtrsim d_{\rm tan}$. In contrast, significant absorption at negative $v_{\rm LSRK}$ in Quadrant I traces Outer Galaxy gas ($R>R_0$) and therefore requires the continuum source to lie beyond that gas.

\begin{figure} 
    \centering
    \includegraphics[width=\linewidth]{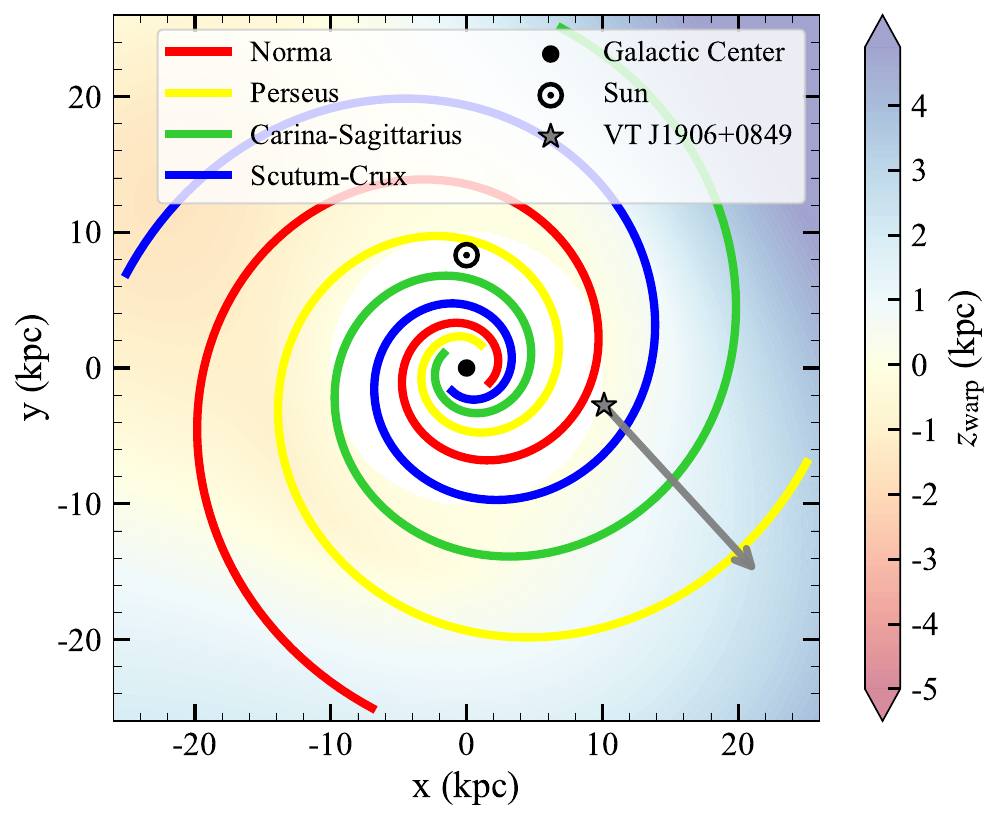}
    \caption{Top-down view of the Milky Way using the spiral arm model of \cite{2008AJ....135.1301V}. The Perseus, Sagittarius-Carina, Scutum-Crux, and Norma-Cygnus arms are shown in yellow, green, blue, and red, respectively. VT J1906+0849 is marked as a gray star, with the gray arrow indicating its estimated distance range. The color map shows the Galactic disk warp for $\phi_1=\phi_2=0^\circ$.}
    \label{fig:top_down}
\end{figure}

\begin{figure}
    \centering
    \includegraphics[width=\columnwidth]{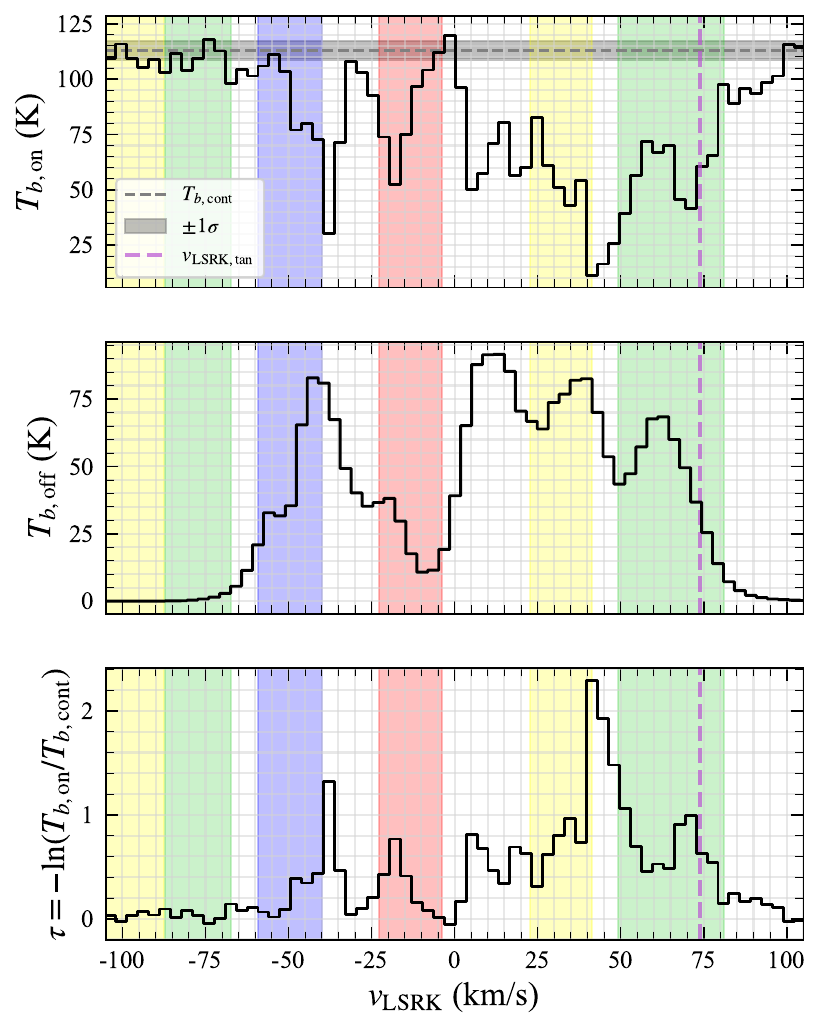}
    \caption{The on-source (top), off-source (middle), and optical depth (bottom) spectra toward VT J1906+0849. The dashed horizontal gray line indicates the continuum level, the shaded gray band represents the $\pm1\sigma$ uncertainty, the dashed vertical purple line indicates $v_{\rm tan}$, and the remaining shaded vertical bands represent spiral arm crossings, with colors matching the arms in Figure \ref{fig:top_down}. 
    }
    \label{fig:hi_spec}
\end{figure}

Using the 2025 VLA HI data, we construct the 
HI spectral line cube, continuum flux, and $1\sigma$ noise using STATCONT \citep{2018A&A...609A.101S}. The on-source absorption spectrum is extracted at the pixel corresponding to the peak continuum emission. Because the VLA data are insensitive to diffuse large-scale HI emission, we use the HI4PI all-sky single-dish HI survey \citep{2016A&A...594A.116H} as an emission reference. The HI4PI spectrum is extracted as a Gaussian beam-weighted average within $1.5\times\theta_{\rm HPBW}\approx24'$ of the source and regridded to match the VLA velocity resolution, $\sim3.3$ km s$^{-1}$.

To associate absorption features with Galactic structure, we adopt the \cite{2008AJ....135.1301V} Milky Way spiral arm model assuming a Sun-Galactic Center distance of $R_0=8.31$ kpc \citep[Figure \ref{fig:top_down};][]{2014ApJ...783..130R}. The spiral arm crossings toward the transient are shown as shaded bands in Figure \ref{fig:hi_spec}, assuming a 9 km s$^{-1}$ velocity dispersion within Quadrant I arms \citep{1995ApJ...448..138M}. We convert between $v_{\rm LSRK}$ and heliocentric distance using the \texttt{kd} package \citep{2018ApJ...856...52W,trey_w_2018_1166001}, which adopts the \cite{2014ApJ...783..130R} rotation curve and $R_0=8.31$ kpc. 

The VLA on-source spectrum, HI4PI emission reference, and inferred optical depth, $\tau=-\ln(S_{\rm on}/S_{\rm cont})$, are shown in Figure \ref{fig:hi_spec}. The spectrum shows strong absorption over much of the inner Galaxy velocity range, from $\approx-25$ to $60$ km s$^{-1}$, with the largest optical depth near $\approx45$ km s$^{-1}$. Absorption extends to velocities consistent with the tangent point, $v_{\rm tan}\approx74$ km s$^{-1}$, implying $d\gtrsim6.2$ kpc. More importantly, distinct absorption features are detected at negative velocities, including features near $\approx-20$ and $-40$ km s$^{-1}$. In the adopted rotation model, the $v_{\rm LSRK}\approx-40$ km s$^{-1}$ feature is associated with Outer Galaxy gas beyond the Solar Circle, placing VT J1906+0849 beyond the Norma-Cygnus arm at $d\gtrsim15$ kpc.

\begin{figure}
    \centering
    \includegraphics[width=\linewidth]{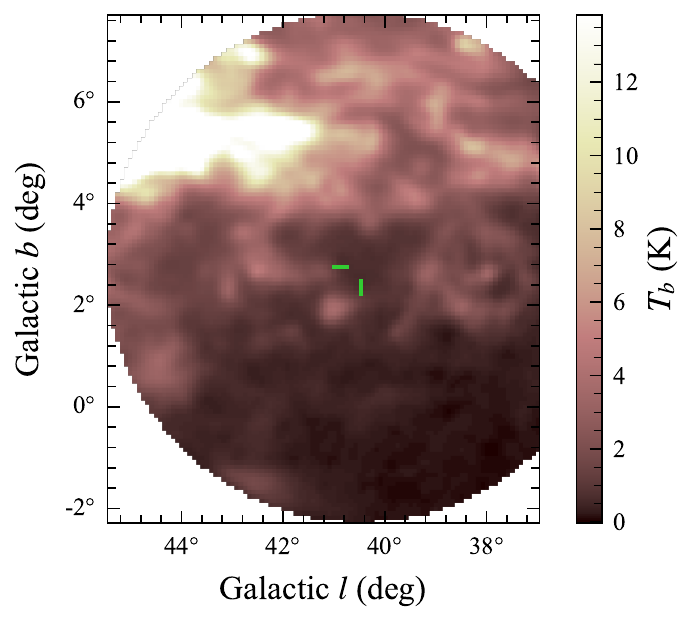}
    \caption{HI4PI $v_{\rm LSRK}\approx-75$ km s$^{-1}$ channel map. No significant HI emission is present at the transient position (marked in green).}
    \label{fig:HI4PI_m75_map}
\end{figure}

\begin{figure}
    \centering
    \includegraphics[width=\linewidth]{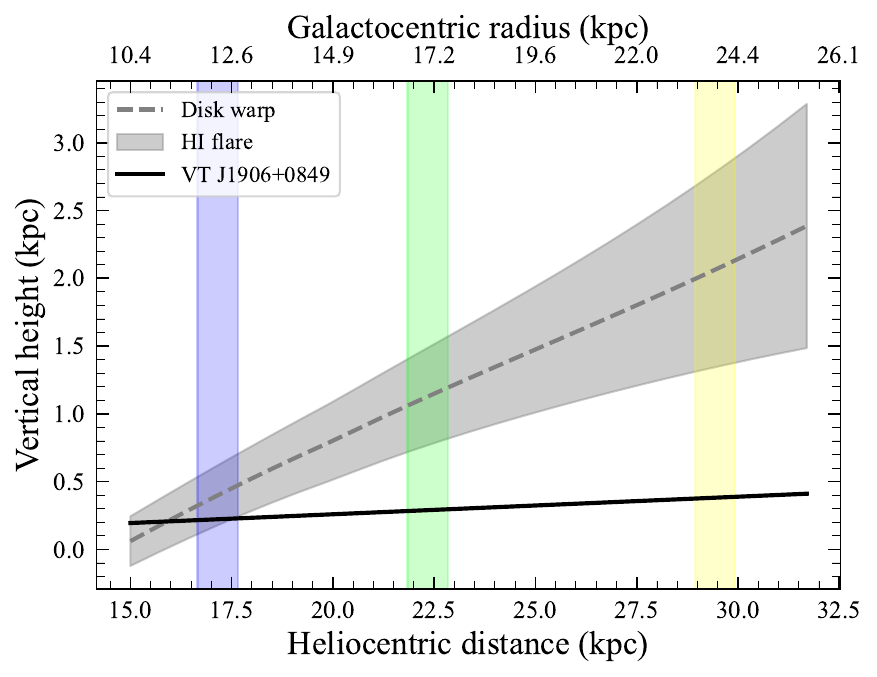}
    \caption{Vertical height of VT J1906+0849 relative to the warped and flared Galactic HI disk. The dashed gray curve shows the \cite{2006ApJ...643..881L} warp model 
    and the shaded gray bands mark the HI flare envelope. The sightline towards VT J1906+0849 is shown in black. Colored vertical bands mark spiral arm crossings using the color scheme of Figure \ref{fig:top_down}.
    }
    \label{fig:hi_scale height_model}
\end{figure}

We do not detect significant absorption at $v_{\rm LSRK}<-40$ km s$^{-1}$. This absence does not provide a comparably strong upper limit, because absorption can only be detected where sufficiently bright foreground HI lies in front of the continuum source. Inspection of the HI4PI channel maps at more negative velocities (e.g., Figure \ref{fig:HI4PI_m75_map}) shows little corresponding HI emission along this sightline. The lack of absorption at these velocities can therefore be explained by the clumpiness of the cold HI distribution \citep{1990ARA&A..28..215D,2003ApJ...582..756K,2009ApJ...693.1250D}, by the warped and flared structure of the outer HI disk \citep[Figure \ref{fig:hi_scale height_model};][]{2009ARA&A..47...27K,2018MNRAS.481L..21P,2019NatAs...3..320C}, or by some combination of both effects. 

The warp and flare are not used as primary HI distance constraints, but they provide a useful check on the plausibility of a Galactic disk origin. Figure \ref{fig:hi_scale height_model} compares the height of the VT J1906+0849 sightline with the \cite{2006ApJ...643..881L} warp model, including the expected flaring in the HI disk \citep{2008A&A...487..951K}. At $d\approx15-17$ kpc, the source lies within $\sim100$ pc of the warped outer disk mid-plane, consistent with an origin in a young or otherwise disk-associated stellar population \citep[e.g.,][]{2016AstL...42....1B,2021AstL...47..534B}. 

An upper limit on distance, assuming a Galactic source, can be motivated by the observed radial extent of the stellar disk. The Galactic stellar disk has been traced out to Galactocentric radii of $R\gtrsim26$ kpc \citep{2018A&A...612L...8L}, corresponding to a heliocentric distance of $d\approx32$ kpc along the VT J1906+0849 sightline. This is not a sharp cutoff, but it provides a reasonable outer bound for a Galactic disk population.

Together, these considerations motivate an allowed Galactic distance range of $d\approx15-32$ kpc. This range is independently supported by proper motion constraints from multi-epoch VLBA detections (Appendix \ref{sec:proper_motion}). Although an extragalactic origin cannot be excluded from these data alone, additional arguments disfavor this possibility (\S\ref{sec:extragal}). If VT J1906+0849 belongs to a population strongly associated with the Galactic disk, a distance near the lower end of this range, $d\approx15-17$ kpc, would be geometrically favored.

\subsection{Constraints from VLBA imaging}

The radio source is marginally resolved in the 2025 VLBA images. Compared with synthesized beams of $4.9\times1.6$ mas in C-band and $4.6\times0.9$ mas in X-band, single elliptical Gaussian fits yield deconvolved major and minor axes of $3.2\pm0.2,1.9\pm0.3$ mas in C-band and $1.4\pm0.2,0.7\pm0.3$ in X-band, respectively (Table \ref{tab:vlba_fits}). The morphology is therefore best described as a compact, slightly elongated Gaussian, with no resolved disk, shell, or jet-like structure.

\subsubsection{Scattering broadening} 

If the source size varies with frequency according to a power law $\theta\propto\nu^{-\kappa}$, the index $\kappa$ can be derived from observations at two frequencies
$$\kappa=-\frac{\ln(\theta_2/\theta_1)}{\ln(\nu_2/\nu_1)}. $$
For angular broadening dominated by interstellar scattering, the expected scaling is $\theta_{\rm sc}\propto\nu^{-2}$, with slightly steeper dependence, $\kappa\approx2.2$, expected for a Kolmogorov spectrum \citep{1990ARA&A..28..561R,2002astro.ph..7156C}. The measured deconvolved source size decreases with increasing frequency, implying $\kappa\approx1.5$ (major axis) to $\kappa\approx1.9$ (minor axis) between 4.9 and 8.4 GHz, broadly consistent with interstellar scattering. NE2025 \citep{2026arXiv260211838O} predicts angular broadening of $\theta_{\rm sc}\sim2.9$ mas (4.9 GHz) and $\sim1.0$ mas (8.4 GHz) for a far-side Galactic source, comparable to the observed sizes, whereas the extragalactic prediction ($\theta_{\rm sc}\sim5.2$ mas and $\sim1.8$ mas) exceeds the measured deconvolved size.

However, the uncertainties in the NE2025 angular broadening predictions are substantial, particularly at low-latitude sightlines such as this one ($b\approx0.74^\circ$). The LOS intersects multiple spiral arms and localized density structures, and small positional offsets ($\sim0.2^\circ$) can produce $\approx20\%$ variations in the predicted Galactic scattering size. More generally, comparisons to pulsar measurements indicate an rms scatter of $\sim0.65$ dex between predicted and observed scattering times \citep{2026arXiv260211838O}, reflecting sensitivity to the location of the dominant scattering screens. Using \texttt{MWprop} \citep{2026arXiv260211838O}, we estimate that scattering along this sightline is dominated by a component at $\sim5$ kpc.

\subsubsection{Brightness temperature} 
Brightness temperature, $T_B$, is the temperature that a blackbody would require to produce the observed radio surface brightness. For a source approximated as an elliptical Gaussian, $$T_B\approx1.38\times10^{12}\left(\frac{S_\nu}{\rm Jy}\right)\left(\frac{\rm mas^2}{\theta_{\rm min}\theta_{\rm maj}}\right)\left(\frac{\rm GHz}{\nu}\right)^{2}\rm\ K $$ \citep{2025ApJS..277...50K}. Using the resolved 2010 and 2022 VLBA measurements reported by \cite{2025ApJS..276...38P}, we estimate $T_b\gtrsim10^9$ K in 2010 and $\gtrsim10^8$ K in 2022. Applying the same relation to our 2025 VLBA observations yields $T_B\gtrsim4.9\times10^8$ K at 4.9 and 8.4 GHz. This likely represents a lower limit on the intrinsic brightness temperature, as the observed size could be inflated by scattering broadening. Such brightness temperatures are well above those expected from thermal emission processes ($\sim10^4$ K), indicating a non-thermal origin.

\subsection{Radio emission mechanisms}

The broadband radio observations of VT J1906+0849 provide insight into the physical mechanism responsible for the emission. In particular, the shape of the radio spectrum can reveal both the underlying radio emission mechanism and the absorption processes responsible for the spectral turnover, while the temporal evolution of the radio light curve provides complementary information on the energetics and variability of the radio transient. In the following subsections, we first model the radio spectra across multiple epochs and then analyze the long-term $5-8$ GHz light curve to characterize the variability timescale of the source, which can aid in source classification.

\subsubsection{Radio spectral fitting}\label{sec:radio_spec_fits}

Most of the radio spectra of VT J1906+0849 between 2017 and 2026 feature a spectral turnover (Figure \ref{fig:single_epoch_multi_freq_radio}). In GHz-peaked spectra, the spectral turnover generally arises from synchrotron self-absorption (SSA) or free-free absorption \citep[FFA;][]{1998PASP..110..493O,2021A&ARv..29....3O}.

We use Bayesian evidence to select the best-fit model. The posterior over the prior volume is 
$Z=\int\int ...\int\mathcal{L}(\theta)\Pi(\theta)d\theta $
where $\mathcal{L}$ is the likelihood function and the number of model parameters determines the dimensionality of the integration. We use 
$$\text{ln}\mathcal{L}(\theta)=-\frac{1}{2}\sum_{n}\bigg[\frac{(S_{\nu,\text{o},n}-S_{\nu,\text{m},n})^2}{\sigma_n^2}+\text{ln}(2\pi\sigma_n^2)\bigg] $$ where for the $n$-th data point, $S_{\nu,\text{o},n}$ and $S_{\nu,\text{m},n}$ are the observed and model flux densities at the data point's frequency $\nu$, and $\sigma_n$ is the uncertainty.

Assuming both models are equally likely to describe the data, the ratios of model evidences can be used to support model selection. If $\Delta\ln(Z)=\ln(Z_2)-\ln(Z_1)$, then by the Jefferys scale, $\Delta\ln(Z)\geq3$ and $\Delta\ln(Z)\approx1-3$ are strong and moderate evidence, respectively, that model 2 is preferred over model 1, and $\Delta\ln(Z)<1$ is inconclusive \cite[e.g.,][]{Kass01061995,2012MNRAS.423L..30S,2015ApJ...809..168C}.

\paragraph{Synchrotron self-absorption} We describe the radio spectra using the generalized smoothly broken power law parametrization of \cite{2002ApJ...568..820G},
$$S_\nu=S_0\left[\left(\frac{\nu}{\nu_0}\right)^{-s\beta_1}+\left(\frac{\nu}{\nu_0}\right)^{-s\beta_2}\right]^{-1/s} $$
where $\beta_1$ and $\beta_2$ are the asymptotic spectral indices below and above the break, respectively, and $s$ controls the sharpness of the transition. The parameters $\nu_0$ and $S_0$ correspond to the frequency and flux density at the intersection of the two asymptotic power law branches, not the spectral turnover. 

We model the radio spectra using nested sampling as implemented in \texttt{dynesty} \citep{2004AIPC..735..395S,10.1214/06-BA127,2009MNRAS.398.1601F,2020MNRAS.493.3132S,k2023}. We require convergence at $d\ln Z\leq0.01$. For the broken power law model, we adopt uniform priors $1\leq S_0\leq500$ mJy, $0.05\leq\nu_0\leq10$ GHz, $0\leq\beta_1\leq2.5$, $-2.5\leq\beta_2\leq0$, and $0.1\leq s\leq3$. 

\begin{figure*} 
    \centering
    \includegraphics[width=\linewidth]{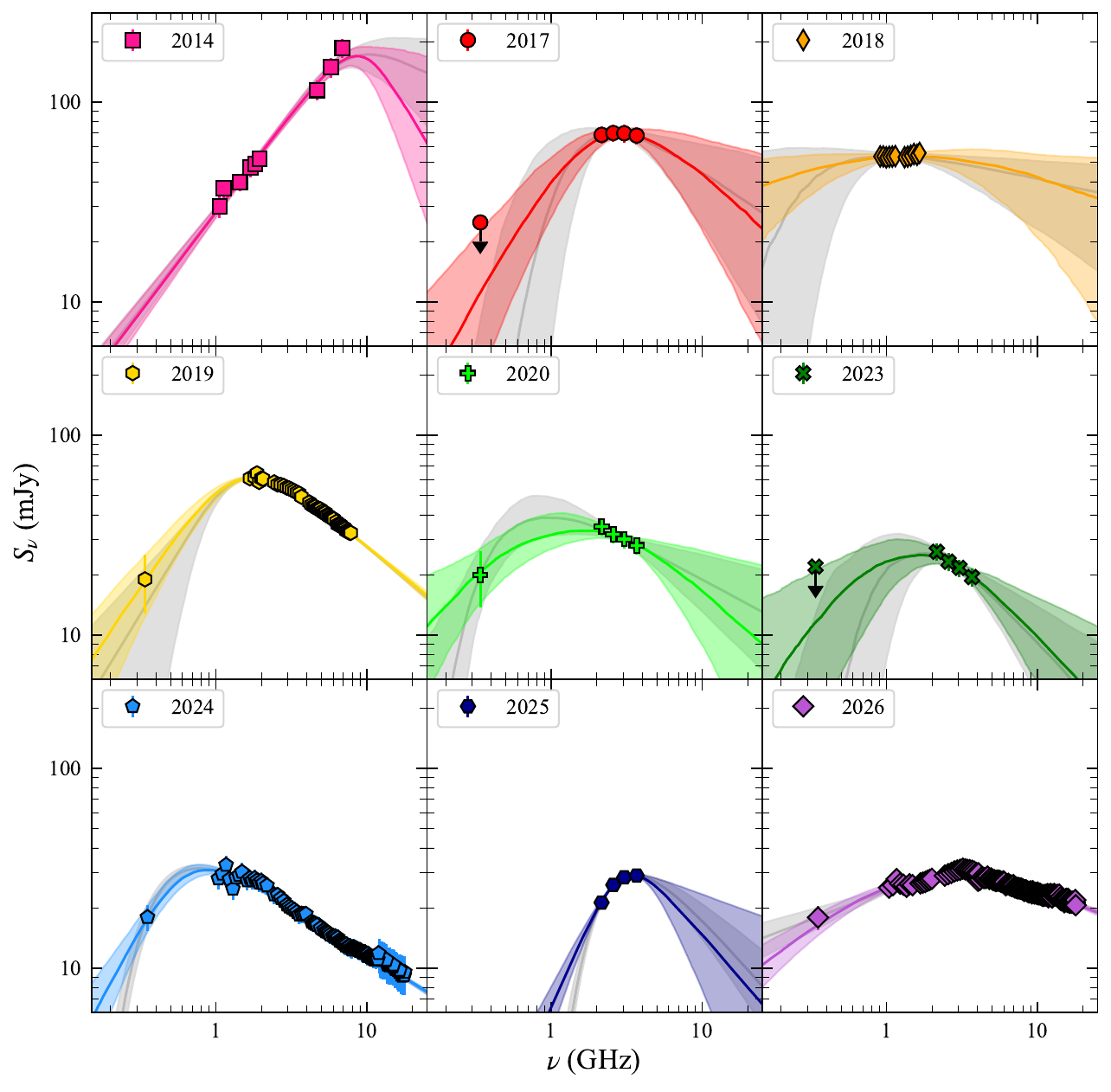} 
    \caption{Quasi-simultaneous broadband radio spectra of VT J1906+0849 across nine epochs from 2014 to 2026. Colored symbols show the measured flux densities for each epoch. The best-fit models for each epoch are shown as a solid curve, while shaded regions indicate the $\pm1\sigma$ posterior envelopes of the model. Colored curves and bands show the synchrotron self-absorption fits, while gray curves and bands show the free-free absorption fits.}
    \label{fig:fitted_radio_spec}
\end{figure*}

The transient does not vary significantly at $\sim340$ MHz over $\lesssim5$ yr timescales (Table \ref{tab:radio_detections}, Figure \ref{fig:single_epoch_multi_freq_radio}). This allows the 2017 October 3 VLITE limit and the 2020 December 27 VLITE detection to be combined with the temporally nearby VLASS epochs (separated by $\sim3$ weeks and $\sim4$ months, respectively) in our spectral modeling. We also include the 2024 July 28 VAST 887.5 MHz detection in our fit to the 2024 August 6 data, as the transient does not evolve significantly between 2024 July 28 to August 15 (Figure \ref{fig:vast_lightcurve}).

The best-fit models and parameters are presented in Figure \ref{fig:fitted_radio_spec} and Table \ref{tab:radio_spec_fit_params}. The optically thick spectral index is best constrained in the 2014, 2019, 2024, and 2026 epochs. None of these epochs are consistent with the canonical value of $\beta_1=2.5$ expected for a homogeneous SSA source with a power law electron energy distribution \citep{1970ranp.book.....P,1979rpa..book.....R}. However, this canonical slope is an idealized limiting case. In an unresolved or inhomogeneous synchrotron source, spatial gradients in magnetic field strength, relativistic electron density, and optical depth can broaden the turnover and produce an effective optically thick spectral index shallower than 2.5 \citep[e.g.,][]{1979ApJ...232...34B,1981ApJ...243..700K}. Similarly, the superposition of multiple self-absorbed components with different turnover frequencies can flatten the integrated spectrum. Thus, the flatter optically thick slopes measured in these epochs do not rule out SSA, but they do disfavor a single homogeneous SSA component and instead point to an inhomogeneous or multi-component emitting region.

The optically thin spectral index constrains the underlying relativistic electron distribution. For optically thin synchrotron emission from a $N(E)\propto E^{-p}$ electron population, $\beta_2=-(p-1)/2$ \citep{1979rpa..book.....R} and $\beta_2=-0.5$ corresponds to $p=2$, the canonical test particle result for diffusive shock acceleration at a strong non-relativistic shock \citep{1983RPPh...46..973D,1987PhR...154....1B}. The best constrained optically thin spectra are in 2019, 2024, and 2026. The 2019 spectrum is consistent with $\beta_2\leq-0.5$, as is the 2024 spectrum when considering uncertainties, but the 2026 spectrum is flatter than this limit. If interpreted as optically thin synchrotron from a single power law electron population, the 2026 spectrum therefore requires an electron distribution harder than the canonical limit, which can be explained by nonlinear shock acceleration, residual opacity, or multiple synchrotron components \citep{1979ApJ...232...34B,2001RPPh...64..429M,2006MNRAS.367.1083K}.

For a passively fading SSA source, the turnover frequency, $\nu_{\rm pk}=\nu_0(-\beta_1/\beta_2)^{1/[s(\beta_1-\beta_2)]}$, is expected to decrease with time as the emitting region expands and becomes optically thin. Instead, we find non-monotonic evolution, with $\nu_{\rm pk}\approx1.9$, $0.8$, and $2.2$ GHz in 2019, 2024, and 2026, respectively, and additional evidence for a high frequency ($\gg4$ GHz) turnover in 2025 from the VLASS Epoch 4 and VLBA detections. This reversal in $\nu_{\rm pk}$ is inconsistent with simple passive expansion and suggests renewed particle injection, evolving opacity or magnetic field conditions, confinement, or multiple emitting components.

The rapidity of this spectral evolution also constrains the physical scale of the region responsible for the changing radio emission. The substantial change in the radio spectra between the 2025 and 2026 detections occurs over a time interval corresponding to a light-crossing scale of $\sim0.14$ pc. If the evolution is intrinsic, causality requires the region in which the opacity and/or synchrotron-emitting conditions are changing to be $\lesssim0.14$ pc in size. This is substantially smaller than the pc-to-kpc radio structures typically associated with unbeamed GHz-peaked spectrum or the $\sim1-20$ kpc structures of compact steep spectrum AGN \citep{1995A&A...302..317F,1998PASP..110..493O,2000MNRAS.319..445S}.


\begin{deluxetable}{rlllllll}
\tablewidth{0pt}
\tablecaption{Best-fit parameters of the smoothed broken power law model fitted to the single-epoch broadband radio spectra. 
\label{tab:radio_spec_fit_params}}
\tablehead{
\colhead{Year} & \colhead{$S_{0}$ (mJy)} &
\colhead{$\nu_{0}$ (GHz)} & \colhead{$\beta_1$} & \colhead{$\beta_2$} & \colhead{$s$} & \colhead{$\ln(Z)_{\rm SSA}$}
}
\startdata
2014 & $240_{-50}^{+200}$ & $9_{-3}^{+1}$ & $1.0_{-0.1}^{+0.4}$ & $-1\pm1$ & $2.3_{-2}^{+0.7}$ & $-37.9\pm0.1$ \\ 
2017 & $100_{-30}^{+300}$ & $2_{-2}^{+4}$ & $1.5\pm1$ & $-0.7_{-2}^{+0.6}$ & $2\pm1$ & $-18.3\pm0.1$ \\ 
2018 & $80_{-20}^{+200}$ & $1_{-1}^{+7}$ & $0.5_{-0.5}^{+2}$ & $-0.4_{-2}^{+0.4}$ & $1.3_{-0.9}^{+2}$ & $-37.9\pm0.1$ \\ 
2019 & $100_{-20}^{+100}$ & $1.4_{-0.7}^{+0.3}$ & $1.1_{-0.4}^{+1}$ & $-0.63_{-0.2}^{+0.06}$ & $1\pm1$ & $-64.0\pm0.1$ \\
2020 & $50_{-20}^{+200}$ & $1_{-1}^{+6}$ & $0.7_{-0.6}^{+2}$ & $-0.6_{-2}^{+0.5}$ & $1\pm1$ & $-18.2\pm0.1$ \\
2023 & $40_{-20}^{+300}$ & $1_{-1}^{+5}$ & $1\pm1$ & $-0.9_{-2}^{+0.7}$ & $1\pm1$ & $-13.6\pm0.1$ \\ 
2024 & $42_{-6}^{+10}$ & $0.6_{-0.2}^{+0.3}$ & $1.4_{-0.7}^{+1}$ & $-0.47\pm0.04$ & $2_{-1}^{+0.9}$ & $-230.3\pm0.1$ \\ 
2025 & $50_{-10}^{+60}$ & $3.4_{-0.9}^{+0.6}$ & $1.7\pm0.7$ & $-1.6_{-0.9}^{+1}$ & $1.4_{-0.9}^{+1}$ & $-13.3\pm0.1$ \\
2026 & $50_{-10}^{+100}$ & $2_{-1}^{+4}$ & $0.7_{-0.4}^{+1}$ & $-0.33_{-0.1}^{+0.5}$ & $10_{-0.8}^{+2}$ & $-278.2\pm0.1$ \\ 
\enddata
\tablecomments{Column 1 lists the year of the observation. Columns 2 and 3 give the flux density and frequency at which the two power laws, described by the spectral indices $\beta_1$ (column 4) and $\beta_2$ (column 5) intersect. Column 6 gives the smoothing parameter, and column 7 gives the corresponding Bayesian evidence, $\ln(Z)$.}
\end{deluxetable}

\paragraph{Free-free absorption} For an external, inhomogeneous free-free absorbing screen, we adopt the parameterization of \cite{1997ApJ...485..112B}, as applied by \cite{2003AJ....126..723T} and \cite{2015ApJ...809..168C}. 
Assuming that the optical depths follow a power law distribution, $\eta^p\propto\int(n_e^2T_e^{-1.35})^pdl$, the observed spectrum is 
$$S_\nu=S_1(p+1)\left(\frac{\nu}{\nu_1}\right)^{2.1(p+1)+\alpha}\gamma\left[p+1,\left(\frac{\nu}{\nu_1}\right)^{-2.1}\right] $$
where $\gamma\left[p+1,(\nu/\nu_1)^{-2.1}\right]$ is the lower incomplete gamma function, $S_1$ is the intrinsic synchrotron flux density normalization at $\nu_1$, $\alpha$ is the intrinsic optically thin spectral index such that $S_\nu\propto\nu^\alpha$ at high frequencies, and $p>-1$ characterizes the distribution of free-free optical depths across the source. 

The parameter $\nu_1$ is defined as the frequency at which the maximum free-free optical depth equals unity. Because different sightlines encounter different optical depths, $\nu_1$ is not generally identical to the spectral turnover frequency.

In our fitting procedure, we adopt log-uniform priors corresponding to $1\leq S_1\leq 500$ mJy, $0.03\leq\nu_1\leq10$ GHz, and uniform priors $-2.5\leq\alpha\leq0$ and $-0.99\leq p\leq10$. 

Figure \ref{fig:fitted_radio_spec} shows the free-free absorption fits. Best-fit parameters and Bayesian evidence for each epoch are presented in Table \ref{tab:logZ_vals}. We find moderate Bayesian evidence that synchrotron self-absorption is preferred in 2019 and 2026, and inconclusive evidence in 2024 (Table \ref{tab:logZ_vals}).

\begin{deluxetable}{rlllllllll} 
\tablewidth{0pt}
\tablecaption{Free-free absorption best-fit parameters and Bayesian evidence comparison between the FFA and SSA models. }\label{tab:logZ_vals}
\tablehead{
\colhead{Year} & \colhead{$S_1$ (mJy)} & \colhead{$\nu_1$ (GHz)} & \colhead{$\alpha$} & \colhead{$p$} & \colhead{$\ln(Z)_{\rm FFA}$} & \colhead{$\Delta\ln(Z)$} & \colhead{Favored model}
}
\startdata
2014 & $200\pm60$ & $9_{-3}^{+1}$ & $-0.5_{-1}^{+0.5}$ & $-0.3_{-0.2}^{+0.7}$  & $-40.1\pm0.1$  & $2.2\pm0.1$ & SSA (moderate) \\
2017 & $120_{-50}^{+60}$ & $1.4_{-0.9}^{+2}$ & $-0.5_{-2}^{+0.5}$ & $5\pm5$ & $-17.8\pm0.1$ & $-0.5\pm0.1$ & Inconclusive \\
2018 & $70_{-20}^{+50}$ & $0.3_{-0.3}^{+2}$ & $-0.2_{-1}^{+0.2}$ & $4\pm5$ & $-38.9\pm0.1$ & $1\pm0.1$  & Inconclusive \\
2019 & $94_{-9}^{+20}$ & $1.5_{-0.5}^{+0.3}$ & $-0.64\pm0.03$ & $-0.1_{-0.3}^{+8}$ & $-65.7\pm0.1$ & $1.7\pm0.1$  & SSA (moderate) \\
2020 & $60_{-30}^{+70}$ & $0.5_{-0.3}^{+3}$ & $-0.4_{-1}^{+0.4}$ & $4_{-4}^{+6}$ & $-19.0\pm0.1$ & $0.8\pm0.1$  & Inconclusive  \\
2023 & $50_{-30}^{+40}$ & $1.0_{-0.8}^{+2}$ & $-0.8_{-1}^{+0.7}$ & $5\pm5$ & $-13.4\pm0.1$ & $-0.2\pm0.1$  & Inconclusive \\
2024 & $53_{-10}^{+5}$ & $0.42_{-0.06}^{+0.3}$ & $-0.48\pm0.03$ & $5\pm4$ & $-230.5\pm0.1$  & $0.2\pm0.1$  & Inconclusive \\
2025 & $60\pm10$ & $2.5_{-0.6}^{+1}$ & $-0.9_{-0.7}^{+0.5}$ & $5\pm4$ & $-12.2\pm0.1$  & $-1.1\pm0.1$ & FFA (moderate) \\
2026 & $35\pm2$ & $2.4_{-0.8}^{+1}$ & $-0.25\pm0.06$ & $0.74_{-0.07}^{+0.1}$ & $-280.3\pm0.1$ & $2.1\pm0.1$ & SSA (moderate) \\
\enddata
\tablecomments{Column 1 gives the year of observation; columns 2-5 give the intrinsic synchrotron flux density normalization at $\nu_1$, the frequency at which $\tau\sim1$, the intrinsic optically thin spectral index $\alpha$, and the normalization factor $p$. Column 6 reports the Bayesian evidence of the FFA fit, column 7 is the difference between the SSA and FFA Bayesian evidence, and column 8 indicates the model favored by Bayesian evidence. Values of $\ln(Z)_{\rm SSA}$ are reported in Table \ref{tab:radio_spec_fit_params}. The 2014 and 2018 spectra show no turnover but are included for completeness.}
\end{deluxetable}




\subsection{Energetics}

Having already constrained a source distance of $d\approx15-32$ kpc (\S\ref{sec:kde}) and strongly disfavoring an extragalactic origin in later sections (\S\ref{sec:extragal}), we estimate the radio and equipartition properties of VT J1906+0849 assuming $d=15$ kpc. The qualitative interpretation is largely insensitive to the adopted distance. These estimates further assume that the GHz spectral turnover is produced by SSA; if free-free absorption instead sets the turnover frequency, the equipartition constraints no longer follow.

The peak observed spectral luminosity of VT J1906+0849 occurred in 2014 at $5\times10^{22}(d/15\rm\ kpc)^2$ erg s$^{-1}$ Hz$^{-1}$ at 6.9 GHz. By integrating the peak flux density in each epoch presented in Table \ref{tab:radio_detections}, we conservatively estimate the total energy radiated over the duration of the outburst ($2005-2026$) as $\gtrsim5\times10^{40}(d/15\rm\ kpc)^2$ erg.

Using the general SSA equipartition methodology commonly applied to spherical compact synchrotron transients and radio supernovae \citep[e.g.,][see also Appendix \ref{sec:ssa_equipartition}]{1977MNRAS.180..539S,1979rpa..book.....R,1994ApJ...426...51R,1998ApJ...499..810C,2013ApJ...772...78B}, we estimate the radius (R), magnetic field (B), energy in electrons ($E_e$), minimum energy ($E_{\rm total}^{(\min)}$), brightness temperature ($T_b$), and cooling time of the synchrotron source ($\tau$) for three epochs with well-constrained spectral peaks and $p\geq2$ (Table \ref{tab:energetics}). 

The minimum total energy at equipartition is $E_{\rm tot}^{(\min)}\gtrsim10^{41}(d/15\rm\ kpc)^{(6p+28)/(2p+13)}$ erg with a radius of $\gtrsim10^{14}(d/15\rm\ kpc)^{(2p+12)/2p+13)}$ cm and a magnetic field strength of $\sim0.35-0.71\times(d/15\rm\ kpc)^{-4/(2p+13)}$ G. $E_{\rm tot}^{(\min)}$ is only a few times larger than the total radio emitted energy to date, which together with the cooling time of a few years, implies extended or continuous energy injection. This is also consistent with the varying optically thin spectral index of the source. 

The inferred equipartition radius changes only modestly between 2019, 2020, and 2023, corresponding to an apparent expansion velocity of $\lesssim10\times (d/15\rm\ kpc)^{(2p+12)/(2p+13)}$ km s$^{-1}$. Such slow evolution is inconsistent with a freely expanding synchrotron source and suggests that changes in the turnover frequency are not driven primarily by expansion. The turnover frequency evolves non-monotonically, decreasing from $\nu_{\rm pk}\approx1.9$ GHz in 2019 to $\approx0.8$ GHz in 2020, rising to $\approx1.1$ GHz in 2023, and shifting to still higher frequencies at later epochs ($\gg4$ GHz in 2025 and $\approx2.2$ GHz in 2026). This behavior points instead to ongoing changes in the emitting region, as discussed in \S\ref{sec:radio_spec_fits}.

\begin{deluxetable}{llll}
\tablewidth{0pt}
\tablecaption{Equipartition properties, assuming SSA is the dominant underlying absorption mechanism. We assume a plasma at distance $d\sim15$ kpc with a ratio of electrons to protons of 1 emits at frequencies $10^7$ to $10^{11}$ Hz in a spherical volume with a volume filling factor of 1. Equipartition calculations are detailed in Appendix \ref{sec:ssa_equipartition}. 
\label{tab:energetics}}
\tablehead{
\colhead{} & \colhead{2019} & \colhead{2020} & \colhead{2023}
}
\startdata
$p$ & 2.26 & 2.2 & 2.8 \\
$R$ (cm) & $1.0\times10^{14}$ & $1.2\times10^{14}$ & $1.4\times10^{14}$\\
$B$ (G) & 0.57 & 0.37 & 0.39 \\
$E_e$ (erg) & $5.0\times10^{40}$ & $3.7\times10^{40}$ & $5.2\times10^{40}$ \\
$E_{\rm total}^{(\min)}$ (erg) & $1.2\times10^{41}$ & $1.8\times10^{41}$ & $7.4\times10^{40}$ \\
$T_b$ (K) & $2.9\times10^{10}$ & $2.9\times10^{10}$  & $2.0\times10^{10}$\\
$\tau$ (yr) & 3 & 7 & 6  \\
\enddata
\tablecomments{In row 1, $p$ is the electron energy index. Rows 2 and 3 are the equipartition radius and magnetic field; rows 4 and 5 are the energy in electrons and minimum equipartition energy; row 6 is the brightness temperature; row 7 is the synchrotron cooling timescale.}
\end{deluxetable}

\subsection{Characterization of the optical+infrared source}\label{sec:nir_spec_analysis}

In this section, we combine optical and near-infrared photometry and spectroscopy, accounting for line of sight reddening, to constrain the origin of the source as Galactic or extragalactic.

\subsubsection{A bright, non-variable multiwavelength counterpart} \label{sec:bright_nonvar_counterpart} 

We compute broadband near-infrared magnitudes by convolving each 1D spectrum with the appropriate UKIRT/UKIDSS filter transmission curve \citep{2020sea..confE.182R}. After interpolating the filter to the spectral grid, we estimate the transmission-weighted mean flux density, $\langle f_\lambda\rangle=\int f_\lambda Td\lambda/\int Td\lambda$, and its uncertainty via trapezoidal integration. Magnitudes are then computed as $m=-2.5\log_{10}(\langle f_\lambda\rangle/f_{0,\lambda})$ where $f_{0,\lambda}$ is the band zeropoint \citep{2007MNRAS.379.1599L}, with uncertainties propagated as $\sigma_m=(2.5/\ln10)(\sigma_{\langle f_\lambda \rangle}/f_\lambda)$.

For the 2021 and 2024 NIRES spectra, we obtain Vega magnitudes of $J=17\pm1$, $H=16.0\pm0.6$, $K=15.1\pm0.3$ and $J=18.7\pm0.5$, $H=17.0\pm0.2$, $K=15.5\pm0.1$, respectively. From the 2025 MOSFIRE spectrum, we find $J=18\pm2$, $H=16.9\pm0.5$, and $K=15.5\pm0.1$ (Table \ref{tab:photometry}). Relative to the 2008 UKIDSS-GPS measurements, our 2021-2025 synthetic photometry shows no evidence for large-amplitude near-infrared variability, with the best constrained $K$ band flux varying by $\ll1$ mag over $\sim17$ yr (Table \ref{tab:photometry}).

We estimate the $izy$ AB magnitudes from the 2023 and 2025 LRIS spectra using the same procedure, adopting the Pan-STARRS transmission curves and zeropoints from \citet{2020sea..confE.182R}. We find $i=22.4\pm0.2$, $z=21.3\pm0.1$, $y=20.7\pm0.1$ in 2023 and $i=22.5\pm0.3$, $z=21.0\pm0.1$, $y=20.0\pm0.1$ in 2025. These measurements are converted to Vega magnitudes in Table \ref{tab:photometry}.

Along this line of sight, the three-dimensional dust map of \cite{2019ApJ...887...93G} implies a minimum extinction of $A_V\gtrsim11.1$ mag at $d\gtrsim15$ kpc, increasing to $A_V\approx14.3$ mag for the total Galactic column \citep{2011ApJ...737..103S}. However, the total column value should not be treated as a formal measurement at the transient's Galactic latitude ($b\approx0.74^\circ$). The nominal \cite{1998ApJ...500..525S} normalization uncertainty is $\sim10\%$ at $|b|>5^\circ$, but the maps are explicitly uncertain for $|b|<5^\circ$ because contaminating sources were not removed. Additionally, empirical comparisons of high-excitation or low-latitude regions indicate systematic errors of order tens of percent \citep{1999ApJ...512L.135A,2021MNRAS.503.5351S}.

We therefore adopt a representative 30\% systematic uncertainty on the total column extinction, corresponding to $\sigma_{A_V}\approx4$ mag. The range $11.1\leq A_V\leq14.3$ mag corresponds to $7.2\leq A_i\leq8.8$, $5.6\leq A_z\leq6.9$, and $4.6\leq A_y\leq5.7$ in the Pan-STARRS $izy$ bands, with systematic uncertainties of approximately $3,$ $2$, and $2$ mag, respectively, for the total column correction.

Assuming a Galactic source at $d\gtrsim15$ kpc, extinction-corrected 2025 optical magnitudes are $i\lesssim15.3$, $z\lesssim15.4$, and $y\lesssim15.4$ (AB), corresponding to absolute magnitudes of $M_i\lesssim-0.6,$ $M_z\lesssim-0.5$, and $M_y\lesssim-0.5$ mag. Adopting $d\approx32$ kpc as a representative maximum distance and $A_V\approx14.3$ mag, a Galactic source would instead have lower limits of $M_i\gtrsim-3.8$, $M_z\gtrsim-3.4$, and $M_y\gtrsim-3.2$ mag.
 
If the source were extragalactic and located behind the full Galactic dust column, applying the same extinction correction would imply intrinsic apparent magnitudes of $i\lesssim13.6, z\lesssim14.1,$ and $y\lesssim14.3$, modulo the systematic uncertainty in the low-latitude extinction correction. This brightness is comparable to that of a nearby supernova at peak luminosity: a normal Type Ia SN with characteristic $M\sim-19$ mag \citep{1993ApJ...413L.105P,2014ApJ...795...44R,2018ApJ...859..101S} would reach $i\sim13.6$ at a distance of $\sim30$ Mpc. However, unlike an ordinary supernova-like optical transient, the counterpart is detected in Pan-STARRS in the early 2010s and again in our 2021, 2023, and 2025 optical follow-up with no evidence of significant ($\gg1$ mag) variability, indicating persistence over more than a decade.

The lack of optical fading instead favors a luminous, perpetual background source, such as a quasar. The brightest quasar on sky, 3C 273 \citep{1963Natur.197.1040S}, lies at $z=0.158$ ($d_L\approx750$ Mpc) and has $i\approx13.6, z\approx12.9,$ and $y\approx12.8$ in PS1 \citep{2016arXiv161205560C}. The dereddened counterpart of VT J1906+0849 would therefore be comparable in apparent brightness to 3C 273. If located at a similar distance, this would require a similarly luminous source. However, because the available cosmological volume increases rapidly with distance, a randomly encountered background quasar is more likely to lie substantially farther away than 3C 273, in which case the same apparent brightness would imply an optical luminosity exceeding that of 3C 273. Thus, a background AGN interpretation requires an unusually luminous optical source.

\subsubsection{An unresolved optical+infrared source}

We test whether the optical and infrared counterpart is resolved in Pan-STARRS $izy$, UKIDSS-GPS $JHK$, Spitzer/GLIMPSE IRAC bands $1-4$, and Spitzer/MIPS 24 $\mu$m images, using the fitting procedure described in \S\ref{sec:opt_archival}. In each band, we compare the target FWHM with the empirical PSF FWHM, defined as the median FWHM of nearby reference stars. We define $Z_{\rm ext}=({\rm FWHM}_{\rm target}-{\rm FWHM}_{\rm PSF})/\sigma$, where $\sigma$ includes both the target FWHM uncertainty and the scatter in the stellar PSF measurements. A source is considered resolved if $Z_{\rm ext}>3$.

The counterpart is unresolved in all optical and infrared images. In Pan-STARRS, $Z_{\rm ext}=-0.2\sigma$, $-0.1\sigma$, and $-1.0\sigma$ in $i$, $z$, and $y$, corresponding to $3\sigma$ limits on the deconvolved FWHM of $<0.8''$, $<0.8''$, and $<1.0''$. The UKIDSS-GPS data provide the strongest constraints. The source is unresolved in $J$ ($-1.2\sigma$) and $H$ ($1.0\sigma$) bands, with deconvolved FWHM limits of $<0.6''$ and $<0.5''$, respectively. In $K$, the target is broader than the median stellar PSF at the $2.4\sigma$ level, corresponding to a nominal deconvolved FWHM of $0.5''$. However, we still treat the source as unresolved in $K$ as this does not meet our $3\sigma$ threshold. At longer wavelengths, $Z_{\rm ext}=-0.4\sigma$, $0.2\sigma$, and $0.7\sigma$ in IRAC bands $1-3$, with FWHM limits of $<2.0''$, $<1.9''$, and $<2.0''$. In IRAC band 4 and MIPS 24 $\mu$m, where the PSF is broader and the background more structured, Gaussian fits remain consistent with the empirical PSF, with limits of $<3.9''$ and $<6.3''$, respectively.

We therefore conclude that the multiwavelength counterpart is unresolved at the resolution of the available optical and infrared imaging. The most stringent constraint comes from UKIDSS-GPS, limiting any resolved component to deconvolved ${\rm FWHM}\lesssim0.5''$, or $8\times10^3$ AU at 15 kpc.

\subsubsection{A variable broad emission line}

\begin{figure*} 
    \centering
    \includegraphics[width=\linewidth]{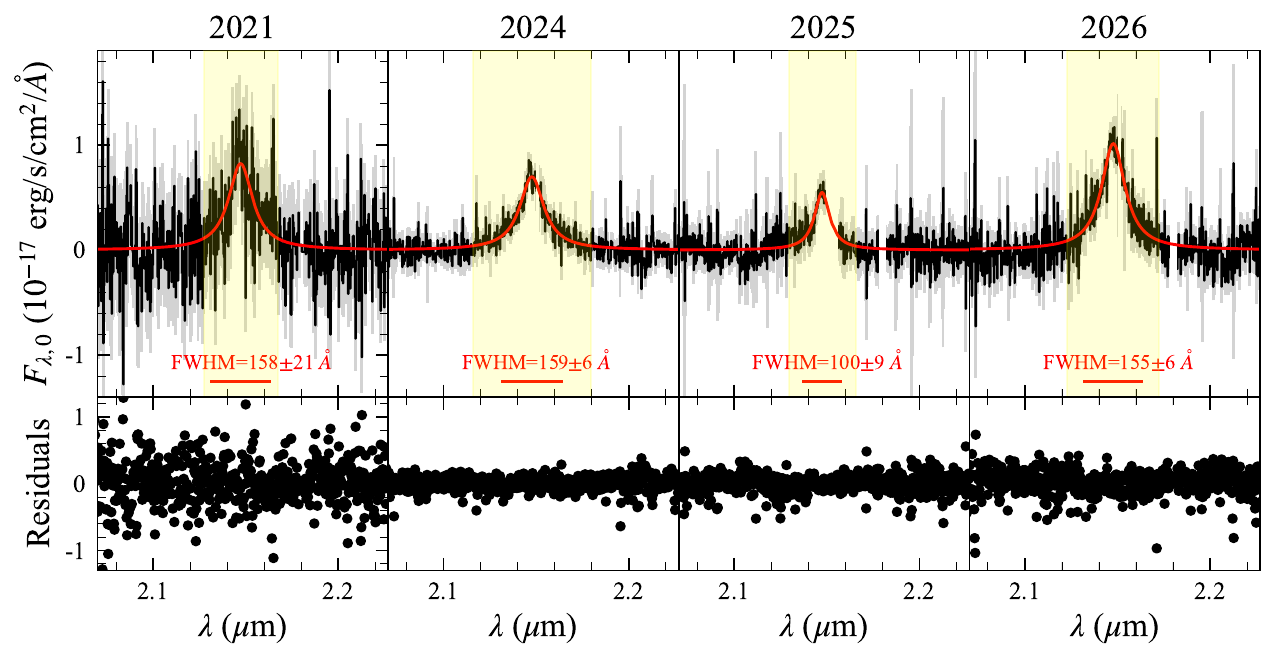}
    \caption{Continuum-subtracted near-infrared line profiles from NIRES (2021, 2024, 2026) and MOSFIRE (2025), together with the best-fit Lorentzian profiles (red curves). The shaded regions indicate the line core. Residuals from the Lorentzian fits are shown in the lower panels.
    }
    \label{fig:nir_fits}
\end{figure*}


\begin{deluxetable}{rlllllll}
\tablewidth{0pt}
\tablecaption{Best-fit Lorentzian model parameters for the near-infrared broad line.
\label{tab:nir_fits}}
\tablehead{
\colhead{Year}  & \colhead{$\lambda_c$ (\AA)}  & \colhead{Amplitude (erg/s/cm$^2$/\AA)} & \colhead{Line flux (erg/s/cm$^2$)} &  \colhead{$\rm FWHM_\lambda$ (\AA)}& \colhead{$v$ (km/s)}
}
\startdata
2021 & $21473\pm7$ & $(8.2\pm0.7)\times10^{-18}$ & $(1.6\pm0.2)\times10^{-15}$ & $158\pm20$ & $2200\pm300$ \\
2024 & $21476\pm2$ & $(7.0\pm0.2)\times10^{-18}$ & $(1.52\pm0.05)\times10^{-15}$ & $159\pm6$ & $2220\pm80$ \\
2025 & $21474\pm3$ & $(5.5\pm0.3)\times10^{-18}$ & $(7.9\pm0.6)\times10^{-16}$ & $100\pm9$ & $1400\pm100$ \\
2026 & $21478\pm2$ & $(1.01\pm0.03)\times10^{-17}$ & $(2.15\pm0.07)\times10^{-15}$ & $155\pm6$ & $2160\pm80$\\
\enddata
\tablecomments{Column 1 is the year of observation; columns 2 and 3 are the centroid and amplitude of the Lorentzian fit; column 4 is the emission line flux from the Lorentzian fit; column 5 is the full-width at half-maximum of the fit; column 6 is the fitted velocity width.}
\end{deluxetable}

An emission feature near $2.146$ $\mu$m is the only spectral line detected and is present in all four near-infrared spectra obtained between 2021 and 2026. We model the feature using the \texttt{specutils} package \citep{specutils_software}. For each spectrum, we subtract a local continuum fit with \texttt{fit\_generic\_continuum} over a window centered on the line, excluding the line core (Figure \ref{fig:nir_fits}, shaded yellow) from the fit.

We fit Gaussian, Lorentzian, and Voigt profiles to the continuum-subtracted line using a Levenberg-Marquardt least-squares fitter, weighting by the per-pixel flux uncertainties. For each model, we compute the $\chi^2$, reduced $\chi^2$, and Bayesian Information Criterion (BIC) to evaluate the relative goodness of fit.

In all epochs, the Lorentzian profile minimized both $\chi^2$ and BIC, indicating it provided the statistically preferred description of the line shape. The Gaussian model was strongly disfavored ($\Delta\rm BIC\gtrsim10$), while the Voigt profile did not yield a significant improvement ($\Delta\rm BIC\approx4$). For simplicity, we only report the model parameters for the best-fit model (Lorentz) in Table \ref{tab:nir_fits} and show the fit in Figure \ref{fig:nir_fits}. 

\begin{figure} 
    \centering
    \includegraphics[width=\linewidth]{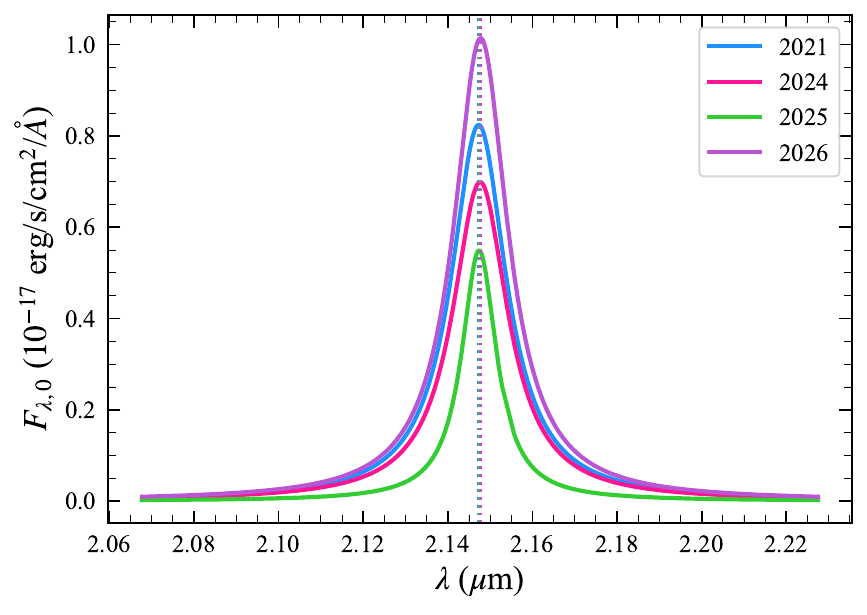}
    \caption{Best-fit Lorentzian models of the continuum-subtracted near-infrared emission feature, with dotted lines marking the fitted line centroids.
    }
    \label{fig:resampled_NIR}
\end{figure}

The line is persistently broad in all epochs. The three NIRES spectra yield mutually consistent velocity widths of $2200\pm300$, $2220\pm80$, and $2160\pm80$ km s$^{-1}$ in 2021, 2024, and 2026, respectively, while the 2025 MOSFIRE spectrum gives a narrower width of $1400\pm100$ km s$^{-1}$. Although the 2026 fit shows mild sensitivity to the adopted fitting region, varying the window by $\pm1000$ km s$^{-1}$ changes the inferred FWHM by only $\sim6\%$, indicating the line remains robustly broad. Additionally, the instrumental velocity resolution of NIRES ($\sim111$ km s$^{-1}$) and MOSFIRE ($\sim83-90$ km s$^{-1}$) differ by only $\sim20-30$ km s$^{-1}$ and amount to a negligible $\lesssim8\%$ of the measured FWHM of the feature.

Figure \ref{fig:resampled_NIR} compares the best-fit Lorentzian profiles across all epochs. The line centroid remains nearly constant over the full observational baseline, while the profile shape and line flux evolve modestly with time. In particular, the 2025 MOSFIRE spectrum exhibits a noticeably narrower profile than the NIRES epochs, whereas the 2021, 2024, and 2026 NIRES observations show remarkably similar widths. 

To test whether the narrower 2025 MOSFIRE width could result from higher noise, we degraded the 2024 NIRES spectrum from its mean line core signal-to-noise ratio of 30 to the 2025 value of 19 and refit the line in 1000 Monte Carlo realizations. The recovered widths remain broad, $2300\pm200$ km s$^{-1}$. This is substantially larger than the 2025 MOSFIRE width, suggesting that the narrower 2025 measurement is not caused solely by fitting the high signal-to-noise core of an intrinsically broad line.

The weighted average centroid of the line, $21476\pm1$ \AA, is coincident with several laboratory atomic transitions in the National Institute of Standards and Technology Atomic Spectra Database \citep{NIST_ASD}, including Fe I at 21475.27 \AA, Xe I at 21475.961 \AA, and Fe II at 21476.3938 \AA. The Fe II transition is the closest astrophysically plausible rest frame match. However, the wavelength coincidence alone is not a unique identification, as a metal line interpretation would generally predict additional Fe II or other atomic features in $K$ band \citep[e.g.,][]{2015MNRAS.447..806D}, which are not detected.

More common nearby $K$ band transitions include Mg II doublet at 2.1375, 2.1438 $\mu$m, He I transitions at 2.1624 and 2.1649 $\mu$m, Br$\gamma$ at 2.1661 $\mu$m, and H$_2$ at 2.1218, 2.1542 $\mu$m \citep[e.g.,][]{1996ApJS..107..281H,1997ApJS..111..445W,2005ApJS..161..154H}.

An isolated, blueshifted Br$\gamma$ line provides the most natural interpretation. The alternative Mg II, He I, and H$_2$ identifications imply distinct line-forming environments: dense UV-irradiated gas for Mg II (ionization energy $E_i\geq7.6,15.0$ eV), a harder radiation field capable of exciting He I ($E_i\geq24.6$ eV), or warm molecular gas with $T_{\rm ex}\sim10^3$ K for near-infrared H$_2$ emission \citep{1987ApJ...322..412B,1996ApJS..107..281H,2006A&A...455..561B}. These scenarios should generally produce the Mg II doublet, additional He I/H recombination lines, or other H$_2$ rovibrational features, none of which are detected.

In contrast, Br$\gamma$ can arise from hydrogen recombination in ionized gas with characteristic temperatures of $T_e\sim5\times10^3-2\times10^4$ K, requiring only an ionizing source capable of maintaining H ionization ($E_i\geq13.6$ eV) and not necessarily strong molecular, metal line, or He I emission. If the feature is Br$\gamma$, the observed centroid implies a blueshift of $\approx2600$ km s$^{-1}$, consistent with emission dominated by approaching material.

Isolated Br$\gamma$ emission, including blueshifted Br$\gamma$ profiles, is observed in embedded young sources where H$_2$ or other nearby lines are absent \citep[e.g.,][]{2001MNRAS.326..524D,2002ApJ...568..771D,2015MNRAS.446.4088T}. We therefore adopt the blueshifted Br$\gamma$ interpretation as the fiducial Galactic line identification, while also considering the possibility of a redshifted line from an extragalactic source in the following section. 

\subsection{An extragalactic interpretation}\label{sec:extragal}

\begin{figure*} 
    \centering
    \includegraphics[width=\linewidth]{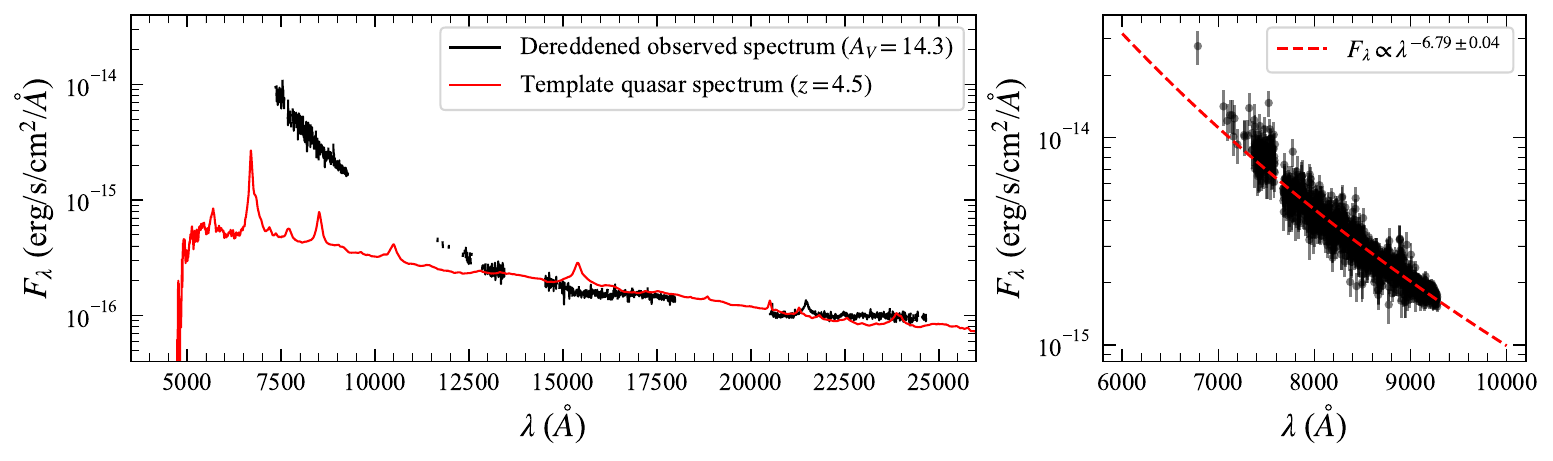}
    \caption{\textit{Left:} Dereddened 2023 LRIS and 2024 NIRES spectra (black) compared to a quasar template spectrum (red) at $z=4.5$. Because only a single broad emission line is present, multiple redshift solutions are possible, and the overplotted template should be regarded as illustrative rather than a meaningful constraint on redshift. \textit{Right: }power law fit to the dereddened optical spectrum (reduced-$\chi^2\approx1.4$).
    }
    \label{fig:sdss_quasars}
\end{figure*}

Among non-transient extragalactic scenarios, only an AGN could plausibly produce such a bright, unresolved source; however, the combination of its radio and optical properties strongly disfavors an AGN interpretation. As discussed in \S\ref{sec:radio_spec_fits}, the significant evolution of the radio spectrum between 2025 and 2026 occurs over only 168 d, corresponding to a light-crossing time of $\sim0.14$ pc for an unbeamed source. Together with the strongly variable, non-monotonic evolution of the spectral turnover, this behavior disfavors an unbeamed peaked-spectrum or compact steep spectrum AGN. Rapid evolution of the convex radio spectra is instead commonly seen in blazars, where a compact, Doppler-boosted, self-absorbed jet can temporarily dominate the radio emission \citep[e.g.,][]{2005A&A...432...31T,2007A&A...475..813O}. Blazars are also strongly overrepresented among variable VLASS sources relative to the underlying compact radio source population \citep{2026OJAp....962407G}. Therefore, if the source is extragalactic, its radio properties favor a jet-dominated, blazar-like AGN.

To assess whether the dereddened 2023 LRIS optical and 2024 NIRES near-infrared spectra are broadly consistent with an AGN origin, we compare them to a composite quasar template constructed from optical spectra of 2200 Sloan Digital Sky Survey (SDSS) quasars and near-infrared spectra of 27 NASA Infrared Telescope Facility quasars, spanning $0.086-3.52\ \mu$m \citep{2001AJ....122..549V, 2006ApJ...640..579G}. The observed and template spectra were resampled onto a common logarithmic wavelength grid, and the template was evaluated over a grid of trial redshifts using a $\chi^2$ minimization. 

Because the spectrum contains only a single broad emission line, multiple quasar emission features can be plausibly associated with the observed line, yielding a wide range of redshift solutions. In this regime, the fit is effectively driven by the continuum shape rather than by secure, multi-line identifications. The best-fit template shown in Figure \ref{fig:sdss_quasars} corresponds to one such solution. 

Significant deviations between the observed spectral energy distribution and the template, particularly in the steep optical continuum, indicate that the source is substantially bluer than a single-component quasar. Fitting a power law to the dereddened optical spectrum gives $\alpha_\lambda\approx-6.79\pm0.04$, where $F_\lambda\propto\lambda^{\alpha_\lambda}$ (Figure \ref{fig:sdss_quasars}, right). This slope is much steeper than typical quasar continua, which generally exhibit $\alpha_\lambda\approx-1.56$ from $1300-5000$ \AA\ and $+0.45$ redward of 5000 \AA\ \citep{2001AJ....122..549V}.

This conclusion is robust to the low-latitude extinction uncertainty discussed in \S\ref{sec:bright_nonvar_counterpart}. Adopting the representative 30\% systematic uncertainty on the total Galactic column and dereddening by the conservative lower limit $A_V\approx10.0$ mag, we still find a steep optical slope of $\alpha_\lambda=-3.42\pm0.05$. 

However, a composite quasar template is not representative of all radio-variable AGN subclasses, and the radio properties of VT J1906+0849 specifically motivate comparison with blazars. Blazars include both flat spectrum radio quasars, which exhibit broad emission lines, and BL Lacs, whose optical emission can be dominated by a nearly featureless synchrotron continuum \citep{1995PASP..107..803U}. The latter provide a particularly relevant comparison for VT J1906+0849. Although BL Lac continua can be blue and approximately power law in form, their optical and ultraviolet slopes are generally substantially shallower than observed here. For example, \cite{1994ApJ...432..547P} find a mean $\alpha_\lambda=-1.12\pm0.42$ for BL Lacs. Even adopting our conservative extinction correction, the slope of VT J1906+0849, $\alpha_\lambda=-3.42\pm0.05$, is considerably bluer. 

The near-infrared emission feature itself does not provide strong evidence for or against an AGN interpretation. Several common broad AGN transitions could be associated with the observed centroid of 21476 \AA. In particular, identification as H$\alpha$ would imply $z\approx2.27$, while H$\beta$ and Mg II would imply $z\approx3.42$ and $z\approx6.67$, respectively. At $z\approx2.27$, the $>200$ mJy detection at $>8$ GHz in 2014 corresponds to a radio spectral luminosity of order $\sim10^{28}$ W Hz$^{-1}$. Such radio powers are high but remain compatible with powerful radio quasars and blazars \citep[e.g., ][]{2019MNRAS.484..204C}.

For an H$\alpha$ identification at $z\approx2.27$, H$\beta$ and [O III] $\lambda5007$ would fall at $\approx1.59$ and $1.64$ $\mu$m, respectively, within the NIRES wavelength coverage, yet no corresponding features are detected. The non-detection of these lines does not rule out a jet-dominated AGN, for which emission lines may appear weak when diluted by the Doppler-boosted non-thermal continuum \citep{1995PASP..107..803U,2011MNRAS.414.2674G}, but precludes a secure H$\alpha$ redshift identification. More generally, we detect no second emission feature that can provide an unambiguous association with any of the candidate AGN transitions.

The line width itself is also not a strong discriminator. Because the identity of the feature is unknown, its FWHM cannot be meaningfully compared with that of a single specific AGN transition or individual quasar. The measured FWHM of $\approx1400-2200$ km s$^{-1}$ lies toward the narrow end of the broad-line AGN population but is not unusual enough to exclude such an interpretation \citep{2006LNP...693...77P,2011ApJS..194...45S}. Broad-line widths depend on the atomic transition, and on source properties and viewing geometry \citep{2017MNRAS.464..385B}, the latter being particularly relevant for a blazar viewed close to the jet axis. We therefore do not regard the line width itself as a strong discriminator between Galactic and extragalactic scenarios.

In conclusion, none of the following independently exclude an AGN origin: the steep slope of the dereddened optical continuum, the wavelength and width of the isolated emission feature, and the varying radio spectra which evolve non-monotonically on $\lesssim1$ yr timescales. Taken together, however, the radio properties restrict a viable extragalactic interpretation primarily to a compact, rapidly evolving, jet-dominated source. As discussed in \S\ref{sec:bright_nonvar_counterpart}, the dereddened counterpart is comparable in apparent brightness to 3C 273, and is likely more luminous. In addition, the optical spectrum is substantially bluer than those of either standard quasars or typical BL Lac objects, and no secure multi-line AGN redshift can be identified. This combination of properties, rather than any individual property, disfavors a blazar-like AGN interpretation for VT J1906+0849.

\subsection{Assessing candidate Galactic progenitors}

With multiple lines of evidence against an extragalactic origin, we now consider Galactic scenarios under the same fundamental assumptions (i.e., a compact, non-variable OIR counterpart). In this framework, distance to the source is $\approx15-32$ kpc, and the line of sight reddening must be $A_V\approx11.1-14.3$ mag. The absolute magnitudes of the optical counterpart are $-3.8\lesssim M_i\lesssim-0.6$, $-3.4\lesssim M_z\lesssim-0.5$, and  $-3.2\lesssim M_y\lesssim-0.5$ mag, corresponding to $L_i\approx(0.1-2.2)\times10^3$ $L_\odot$, $L_z\approx(0.1-1.5)\times10^3$ $L_\odot$, and $L_y\approx(0.1-1.2)\times10^3$ $L_\odot$. Any viable Galactic interpretation must therefore simultaneously reproduce these intrinsic luminosities, the broadband spectral energy distribution (SED), and the observed variable spectral line feature while remaining consistent with the allowed range of extinction.

\subsubsection{A stellar optical counterpart}\label{sec:subsub_opt_counterpart}

\begin{figure*}
    \centering
    \includegraphics[width=\linewidth]{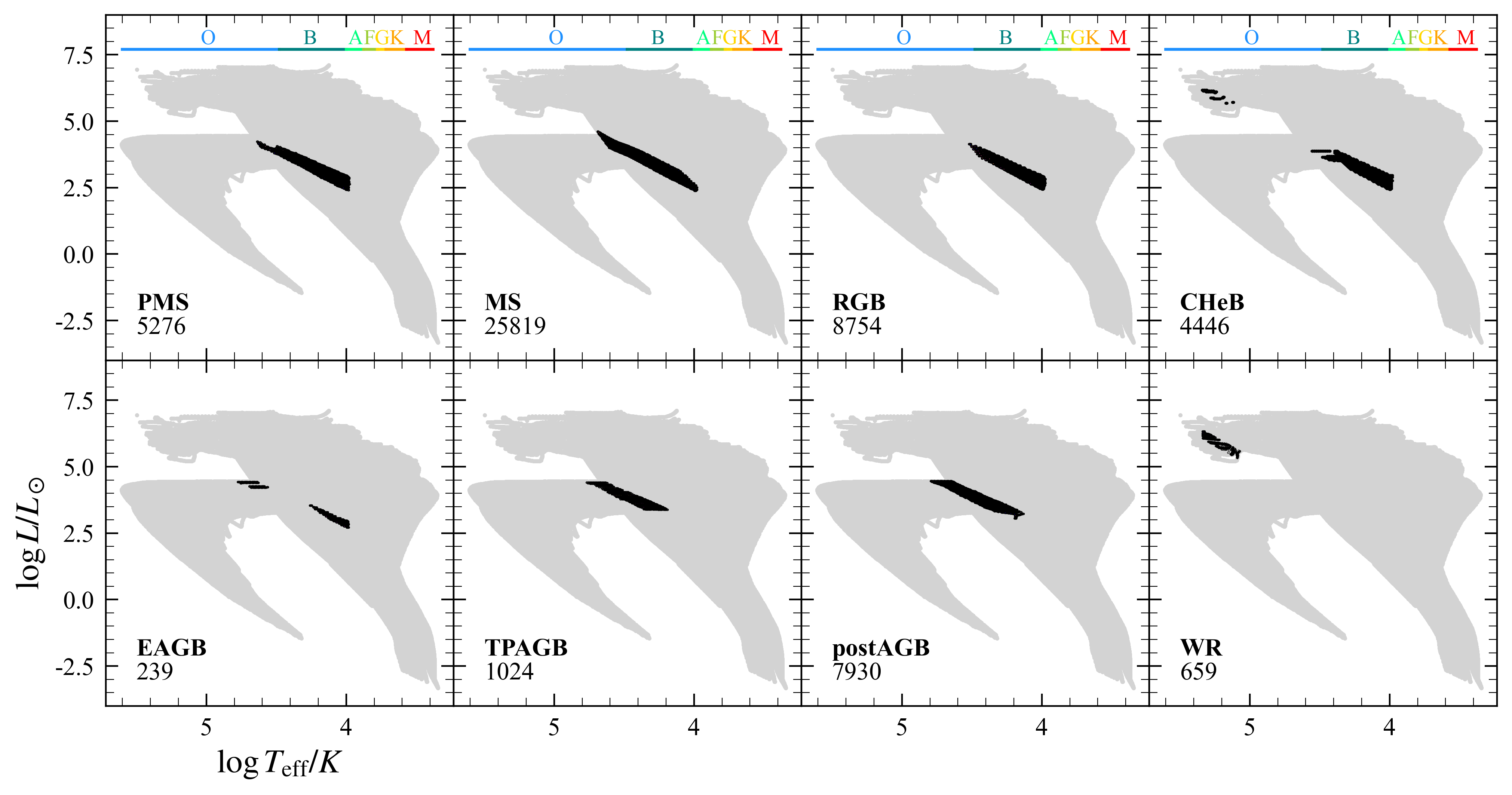}
    \caption{Allowed stellar evolutionary phases for the optical counterpart in the Hertzsprung-Russell diagram. Gray shading shows the full set of MIST stellar models, while black regions indicate models satisfying the criteria listed in \S\ref{sec:subsub_opt_counterpart}. 
    Panels are separated by MIST evolutionary phase: pre-main sequence (PMS), main sequence (MS), red giant branch (RGB), core helium burning (CHeB), early asymptotic giant branch (EAGB), thermally pulsating AGB (TPAGB), post-AGB, and Wolf-Rayet (WR). The number of surviving models in each phase is listed in each panel. 
    }
    \label{fig:HRD}
\end{figure*}

The inferred optical magnitude, together with the compact and non-variable nature of the counterpart, strongly suggests that the optical emission is dominated by a stellar source. We therefore examine whether a single stellar photosphere can reproduce the observed broadband properties, considering both main-sequence and evolved stellar interpretations.

\paragraph{Hertzsprung-Russell  diagram}
To constrain the nature of the optical counterpart, we compared the dereddened photometry to the MESA Isochrones and Stellar Tracks (MIST) synthetic photometry grids \citep{2016ApJS..222....8D,2016ApJ...823..102C}, adopting models at all available metallicities with $v/v_{\rm crit}=0.0$. These grids provide stellar ages, evolutionary states, masses, radii, surface gravity, effective temperatures, bolometric luminosities, and synthetic absolute magnitudes in the Pan-STARRS filters at $A_V=0.0$ mag.

We selected models consistent with the allowed absolute magnitude ranges derived above and additionally required intrinsic colors of $-0.4\lesssim i-z\lesssim0.1$, $-0.2\lesssim z-y\lesssim0.0$, and $-0.6\lesssim i-y\lesssim -0.1$ mag. To remain consistent with non-detection of the source in the bluer Pan-STARRS bands, we further required that the apparent $g$ and $r$ band magnitudes at $d=15$ kpc and $A_V=11.1$ mag fall below the Pan-STARRS $3\pi$ $5\sigma$ limits ($g\approx23.3$, $r\approx23.2$ mag).

The remaining models span several evolutionary states (Figure \ref{fig:HRD}; Table \ref{tab:hrd}), demonstrating that broadband photometry alone does not uniquely determine the stellar class. However, all allowed solutions occupy relatively hot ($T_{\rm eff}\gtrsim10^4$ K) and luminous ($L\gtrsim240 L_\odot$) regions of the Hertzsprung-Russell diagram. In particular, the coolest remaining models have $T_{\rm eff}\approx9400$ K, corresponding to an A0-A2 star.

\begin{deluxetable}{lllllll}
\tablewidth{0pt}
\tablecaption{
For each evolutionary phase shown in Figure \ref{fig:HRD}, we list the range of allowed model ages ($\log\tau$), radii ($R$), masses ($M$), effective temperatures ($T_{\rm eff}$), bolometric luminosities ($L$), and surface gravities ($g$).
\label{tab:hrd}}
\tablehead{\colhead{Evolutionary state} & \colhead{$\log\tau$ (yr)} & \colhead{$R$ ($R_\odot$)} & \colhead{$M$ ($M_\odot$)} & \colhead{$T_{\rm eff}$ ($10^4$ K)} & \colhead{$L$ ($10^3$ $L_\odot$)} & \colhead{$\log g$ (cm s$^{-2}$)}} 
\startdata
PMS & $5.0 - 6.1$ & $2.1 - 10.2$ & $3.6 - 13.8$ & $0.95-4.3$ & $0.25-16.9$ & $3.1 - 4.9$ \\
MS & $5.0 - 8.5$ & $2.1 - 7.6$ & $3.3 - 19.4$ & $0.95 - 4.9$ & $0.24-40.1$ & $3.3 - 4.9$ \\
RGB & $7.4 - 8.6$ & $3.0 - 10.2$ & $2.4 - 9.2$ & $0.94 - 3.3$ & $0.27-13.6$ & $3.0 - 4.3$ \\
CHeB & $6.5 - 10.1$ & $0.9 - 10.6$ & $0.5 - 43.5$ & $0.96 - 22$ & $0.27-1500$ & $2.7 - 6.1$ \\
EAGB & $7.9 - 8.7$ & $1.5 - 10.4$ & $0.8 - 4.8$ & $0.96 - 5.9$ & $0.51-26.4$ & $2.9 - 4.1$ \\
TPAGB & $7.9 - 10.2$ & $1.6 - 6.8$ & $0.5 - 1.0$ & $1.6 - 5.8$ & $2.38-25.2$ & $2.5 - 4.0$ \\
post-AGB & $7.8 - 10.3$ & $1.5 - 7.4$ & $0.5 - 1.1$ & $1.4 - 6.2$ & $1.1-28.5$ & $2.4 - 4.1$ \\
WR & $6.4 - 6.7$ & $0.8 - 1.5$ & $10.8 - 41.4$ & $12 - 22$ & $210-2100$ & $5.3 - 6.1$ \\
\enddata
\end{deluxetable}

\paragraph{Stellar templates}
As an independent test of a stellar interpretation, we compared the observed optical and near-infrared SED to empirical stellar templates from the Pickles Atlas \citep{1998PASP..110..863P}, which provides 131 flux-calibrated stellar spectra spanning $1150-25000$ \AA\ with a sampling interval of 5 \AA\ per pixel and a spectral resolution of 500 \AA. Spectral and luminosity classes include O5-M6 V, B2-K2 IV, O8-M10 III, B2-M3 II, and B0-M2 I. Templates hotter than F-type stars are at solar metallicity; those cooler have a range of metallicities. Pre-main sequence templates are not included.

Motivated by the HR diagram constraints above, which require a hot photosphere $(T_{\rm eff}\gtrsim9400$ K), we restrict our analysis to O-, B-, and early A-type templates.

Each template was reddened using the \cite{1999PASP..111...63F} extinction law with $R_V=3.1$, adopting a grid of visual extinctions spanning $A_V=11.1-14.3$ mag in steps of 0.1 mag. We performed $\chi^2$ minimization against the 2025 LRIS spectrum, fitting the optical data alone to isolate the stellar photospheric component.

The preferred solution is a B3 V template with $A_V\approx11.3$ mag and $T_{\rm eff}\approx19,000$ K. However, this fit is not unique: several templates provide statistically comparable fits ($\Delta\chi^2<10$), including B0 V, B1 I, B3 I, B3 III, and O8 III. Hotter templates prefer larger extinctions, reflecting effective temperature-extinction degeneracy.

\begin{figure}
    \centering
    \includegraphics[width=\linewidth]{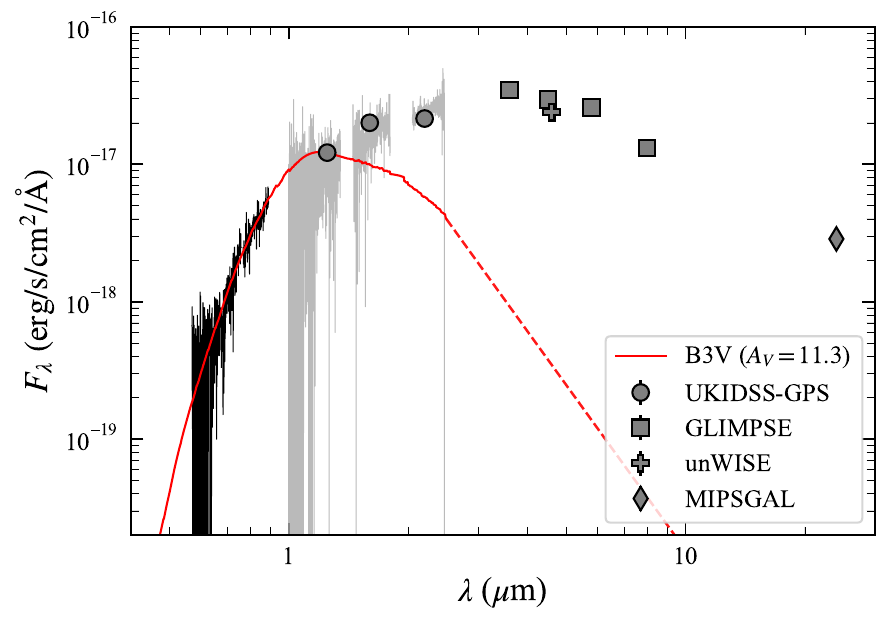}
    \caption{The optical to mid-infrared SED of VT J1906+0849, showing the 2025 optical (black) and 2024 near-infrared (gray) spectra, archival photometry, best-fit reddened stellar template (solid red curve), and Rayleigh-Jeans extrapolation (dashed red line). }
    \label{fig:fitted_spectrum}
\end{figure}

An isolated stellar model underpredicts the observed $H$ and $K$ band fluxes (Figure \ref{fig:fitted_spectrum}). Extrapolation to longer wavelengths using the Rayleigh-Jeans approximation ($F_\lambda\propto\lambda^{-4}$) shows that the GLIMPSE, unWISE, and MIPSGAL photometry lie well above the expected photospheric emission, indicating a strong mid-infrared excess.

We therefore conclude that no single reddened stellar photosphere can simultaneously reproduce the optical spectrum and infrared photometry. Instead, the data favor a hot stellar source dominating the optical emission together with an additional infrared component that becomes increasingly dominant at longer wavelengths.

\paragraph{SEDFitter} To model the infrared excess, we fit the broadband SED using \texttt{sedfitter} \citep{2007ApJS..169..328R,2017A&A...600A..11R}, considering configurations in which a central stellar source is embedded in circumstellar dust in either a disk- or envelope-dominated geometry.

We adopt the ``sp-hmi'' and ``s-p-hmi'' model sets, corresponding to passively irradiated disks with variable inner radii and spherical envelopes with power law density profiles and variable inner cavity radii, respectively. Although these models are designed for young stellar objects, they provide a useful framework for testing whether a stellar central source embedded in dust can reproduce the observed SED.

Models are evaluated via $\chi^2$ minimization, imposing the same extinction and distance constraints as above ($A_V\approx11.1-14.3$, $d\gtrsim15$ kpc), with both treated as free parameters. Only a single envelope model is consistent with the data, while three disk models provide comparably good fits ($\Delta\chi^2<10$).

\begin{figure} 
    \centering
    \includegraphics[width=\linewidth]{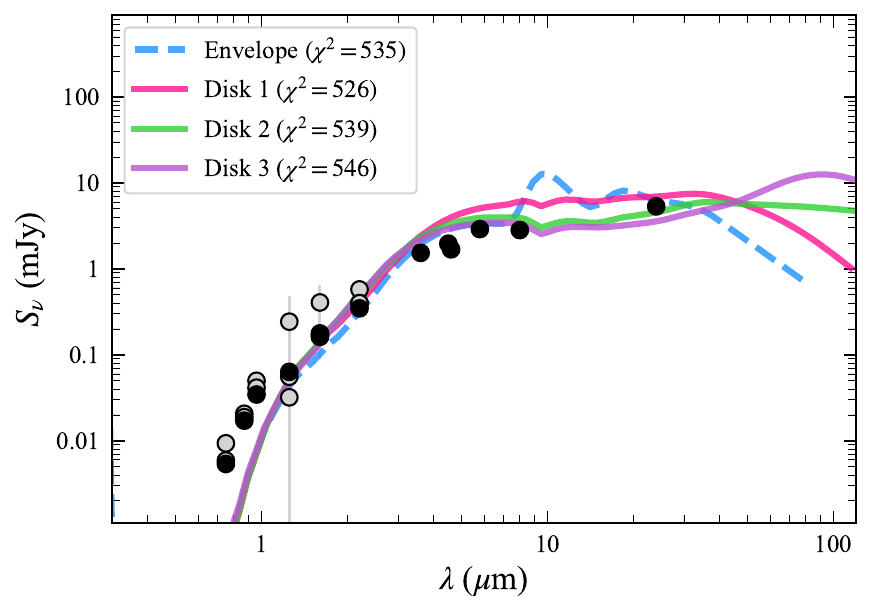}
    \caption{Best-fit \texttt{sedfitter} models for VT J1906+0849. The envelope (blue dashed line) and disk (solid pink, green, and purple lines) models are compared to the optical-infrared SED; black points indicate data used in the fit, while gray points show additional photometry.
    }
    \label{fig:sedfitter}
\end{figure}

All acceptable solutions favor a distant, highly reddened ($A_V\approx11.1$ mag) system hosting a hot, luminous central star. The best-fitting envelope model yields $T_{\rm eff}\approx13400$ K, $R\approx6 R_\odot$, and $L\approx1040 L_\odot$ at a distance of 19.3 kpc surrounded by a dusty envelope with density profile $\rho(r)=3\times10^{-22}(r/1000\rm\ AU)^{-1.543}\mathrm{\ g \ cm^{-3}}$ and an inner radius of $\approx3$ AU. 

The three viable disk models place the source at $d\approx15-18$ kpc and require a central star with $L\approx350-970$ $L_\odot$, $T_{\rm eff}\approx10800-12700$ K, and $R_\star\approx5.4-7.8$ $R_\odot$. The surrounding dusty disks extend from a few AU to $\sim10^2-10^3$ AU with inclinations of $40^\circ-83^\circ$ and masses of $M_d\approx2-260\ M_\oplus$. 

These fits should be interpreted as phenomenological rather than physical. The adopted model grids are not tailored to this source and are subject to degeneracies between extinction, distance, and geometry, so the inferred disk and envelope parameters are not unique. Instead, the \texttt{sedfitter} analysis demonstrates that a hot stellar source embedded in significant quantities of dust can reproduce the observed SED.
 
\subsection{Constraints on soft X-ray luminosity}

For the X-ray non-detection, we assume an absorbed power law spectrum with a conservative photon index $\Gamma=2$. This is appropriate for a hard state accreting black hole (\S\ref{sec:compact_object_acc}), where X-ray emission is dominated by a hard power law component with typical photon indices of $\Gamma=1.5-2.1$ \citep{2006ARA&A..44...49R}. We adopt a Galactic hydrogen column density of $N_H=1.32\times10^{22}$ cm$^{-2}$ \citep{2016A&A...594A.116H}. Using WebPIMMS\footnote{\href{https://heasarc.gsfc.nasa.gov/cgi-bin/Tools/w3pimms/w3pimms.pl}{Link to WebPIMMS}}, this gives a $3\sigma$ absorbed $0.3-10$ keV flux limit of $F_X\leq1.7\times10^{-13}$ erg s$^{-1}$ cm$^{-2}$, corresponding to an unabsorbed flux limit of $F_{X,\rm unabs}\leq3.4\times10^{-13}$ erg s$^{-1}$ cm$^{-2}$. At a distance of 15 kpc, the unabsorbed flux limit corresponds to an intrinsic X-ray luminosity of $L_X\leq9\times10^{33}$ erg s$^{-1}$.

\section{Discussion}\label{sec:discussion}

\subsection{Constraining the underlying physical nature of the radio transient}

Before considering possible interpretations, we summarize the fundamental observational properties of VT J1906+0849:

\begin{itemize}
    \item A decades-long radio transient that reached a peak spectral luminosity of $\approx5\times10^{22}$ erg s$^{-1}$ Hz$^{-1}$ at 6.9 GHz and radiated $\gtrsim5\times10^{40}$ erg in the radio over $\sim20$ yr
    \item A compact size (a few-tens of AU) with $T_b\gtrsim10^8$ K (VLBI) to $T_b\gtrsim10^{10}$ K (equipartition)

    \item An evolving, peaked radio spectrum: the turnover frequency changes non-monotonically with time, moving from $\approx1.9$ GHz in 2019 to 0.8 GHz in 2024 and back to 2.2 GHz in 2026; the high-frequency spectral slope also evolves, flattening from $\approx-0.63$ in 2019 to $\approx-0.33$ in 2026.
        
    \item Location in the warped Galactic disk at $15\lesssim d\lesssim32$ kpc
    \item A broad near-infrared line with a constant centroid position, varying FWHM ($v\approx1400$ to $2220$ km s$^{-1}$), and varying line flux (factor of $\sim3$ fluctuation in $F_{\rm line}$)
    \item An optical-infrared SED consistent with a hot ($T_{\rm eff}\gtrsim10^4$ K) stellar source embedded in a dusty disk or envelope, with an infrared luminosity at most a couple of tens of percent ($\lesssim15\%$) of the stellar bolometric luminosity
    \item No detected X-ray counterpart ($L_X\lesssim9\times10^{33}$ erg s$^{-1}$)
\end{itemize}

To the authors' knowledge, this combination of properties - decades-long transient radio emission, evolving synchrotron spectrum, AU-scale compactness, a high-velocity varying infrared line, a dusty counterpart, and an X-ray non-detection - is unique among previously known sources.

The radio luminosity alone rules out several known classes of Galactic synchrotron transients. The source is $\sim10^9\times$ more luminous than flare stars \citep{2005A&A...436..241S,2017MNRAS.471.3788P}, $\sim10^6\times$ that of magnetically active binaries (RS CVn, Algols; \citealt{1987MNRAS.229..659S,2005ESASP.560..407A,2011ApJ...737..104P,2017MNRAS.471.3788P}) and non-magnetic cataclysmic variables \citep{2008Sci...320.1318K,2015MNRAS.451.3801C}, and $\sim10^5\times$ more than magnetic CVs. 

Although novae can produce compact, non-thermal synchrotron radio emission \citep{2014Natur.514..339C,2016MNRAS.457..887W,2021ApJS..257...49C}, VT J1906+0849 is difficult to reconcile with a nova origin. Its peak radio luminosity exceeds that of typical novae at $5-8$ GHz, and its radio emission has persisted for decades rather than days-to-years timescales observed for nova shocks \citep{2016MNRAS.457..887W,2021ApJS..257...49C}. The multiwavelength counterpart also lacks the rich emission line spectrum expected from nova ejecta \citep{1992AJ....104..725W}. The high brightness temperature therefore supports a non-thermal origin, but the luminosity, duration, and absence of nova-like emission lines argue against a nova interpretation.

Given the restrictive energy requirements, we consider two energy reservoirs capable of powering the radio emission: spin-down luminosity from a pulsar and accretion onto a compact object. 

\subsubsection{Pulsar-powered scenario}
One potential energy reservoir is the rotational energy of a young neutron star. The canonical example is the Crab pulsar, whose spin-down luminosity of $\dot{E}\sim5\times10^{38}$ erg s$^{-1}$ \citep{1974MNRAS.167....1R,2006ARA&A..44...17G,2008ARA&A..46..127H} powers the synchrotron-emitting Crab Nebula through the continuous injection of relativistic particles and magnetic fields. Over its $\sim10^3$ yr lifetime, the Crab has injected $\sim10^{49}$ erg into the surrounding nebula, demonstrating that pulsar spin-down can sustain luminous non-thermal emission for centuries without requiring ongoing accretion. More generally, pulsar wind nebula (PWN) provide a well-established mechanism for converting neutron star rotational energy into long-lived radio synchrotron emission \citep{1984ApJ...278..630R,2006ARA&A..44...17G,2017hsn..book.2159S}.

VT J1906+0849 shares several qualitative similarities with PWN, including a non-thermal radio spectrum, high brightness temperature, and a total radio luminosity that can be accommodated within the spin-down power of a young pulsar, $\dot{E}=4\pi^2I\dot{P}/P^3$, where $I$ is the neutron star moment of inertia, $P$ is the spin period, and $\dot{P}$ is its time derivative. However, unlike the Crab Nebula and other well-studied PWN \citep[e.g.,][]{2001ApJ...556..380H,2002ApJ...568..226M,2004ApJ...616..403S,2008ARA&A..46..127H,2023ApJ...948..119D}, VT J1906+0849 exhibits substantial evolution on human timescales, including a recent radio rebrightening and a shift in the spectral turnover toward higher frequencies. Such behavior implies changes in the particle population, magnetic field strength, or absorbing environment, whereas classical PWN generally evolve gradually on timescales of centuries to millennia \citep{2006ARA&A..44...17G}. Taken together, the observed radio evolution is inconsistent with the gradual expansion and fading expected for a classical PWN.

\subsubsection{Accretion onto a compact object} \label{sec:compact_object_acc}

Accretion onto a compact object naturally explains the evolving synchrotron self-absorbed spectrum, compact emitting region, $T_b\gtrsim10^8$ K, and varying broad emission feature. We explore two potential configurations to explain the radio transient: a ``typical" X-ray binary and a jet-wind interaction.

\paragraph{X-ray binary}
The presence of a B-type or hotter stellar companion suggests a possible high-mass X-ray binary (HMXB configuration). At a distance of 15 kpc, VT J1906+0849 lies within $\approx100$ pc of the warped Galactic disk, consistent with the scale height of HMXBs \citep[$\lesssim150$ pc;][]{2002A&A...391..923G}.

The total energy in particles and magnetic field in VT J1906+0849, $E_{\rm tot}^{(\rm min)}\gtrsim10^{41}$ erg, exceeds that of typical steady compact jets \citep[$\sim10^{36}-10^{38}$ erg;][]{2006csxs.book..381F}, but falls within the range observed for discrete ejection events \citep[$\sim10^{39}-10^{42}$ erg;][]{1999MNRAS.304..865F,2013MNRAS.432..931B} and remains below the most extreme relativistic outbursts \citep[$\gtrsim10^{43}$ erg;][]{1999MNRAS.304..865F,2007MNRAS.378.1111B,2013MNRAS.432..931B}. However, despite this similar energy scale, VT J1906+0849 is difficult to explain as a single discrete ejection because the radio emission is detectable for decades, far longer than the inferred synchrotron cooling time of $<10$ yr (Table \ref{tab:energetics}). While repeated discrete ejections cannot be excluded, the lack of several distinct  flares in the radio lightcurve and the absence of identifiable, outward moving VLBI components favors a long-lived or continuously replenished emitting region.

We now consider the jet power associated with the radio emission. Assuming a steady, flat spectrum, optically thick jet, the 2019 8.6 GHz luminosity implies $P_{\rm jet}\gtrsim7\times10^{38}$ erg s$^{-1}$ \citep[][]{2006MNRAS.369.1451K}. If the accreting compact object is a black hole, the required accretion rate to produce $P_{\rm jet}$ is $\dot{M}_{\rm BH}\gtrsim3\times10^{-7}$ $M_\odot$ yr$^{-1}$ \citep{2006MNRAS.369.1451K}, corresponding to super-Eddington accretion for $M_{\rm BH}\lesssim15(\eta /0.1)^{-1}M_\odot$ where $\eta$ is the accretion efficiency. Sustaining $\dot{M}_{\rm BH}$ for $\sim20$ yr implies a total accreted mass of $\sim10^{28}$ g and a total jet energy release of order $\sim10^{47}$ erg. 

This energy budget is comparable to the energy injected over the same interval by the persistent $L_k\sim10^{39}$ erg s$^{-1}$ baryonic jets of SS 433 \citep{2004ASPRv..12....1F}. The inferred instantaneous jet power is also comparable to the $\gtrsim10^{38}$ erg s$^{-1}$ powers inferred during repeated ejection episodes of GRS 1915+105 \citep{1999MNRAS.304..865F,2000MNRAS.318L...1F}, but substantially exceeds the $\sim10^{36}-10^{37}$ erg s$^{-1}$ time-averaged power inferred for the persistent hard state jet of Cygnus X-1 \citep{2005Natur.436..819G,2007MNRAS.376.1341R}. VT J1906+0849 would therefore lie at the most energetic end of Galactic X-ray binary jets, particularly if its elevated jet power has been maintained for decades.

The inferred FWHM velocity of the blueshifted Br$\gamma$ emission line is consistent with the low-ionization optical and near-infrared disk winds observed in X-ray binaries \citep[see][]{2026SSRv..222...39M} and suggests that the line-emitting gas is part of a fast outflow. The strong net blueshift may indicate an asymmetric geometry in which the approaching side is preferentially visible while emission from the receding side is obscured.

\begin{figure}
    \centering
    \includegraphics[width=\linewidth]{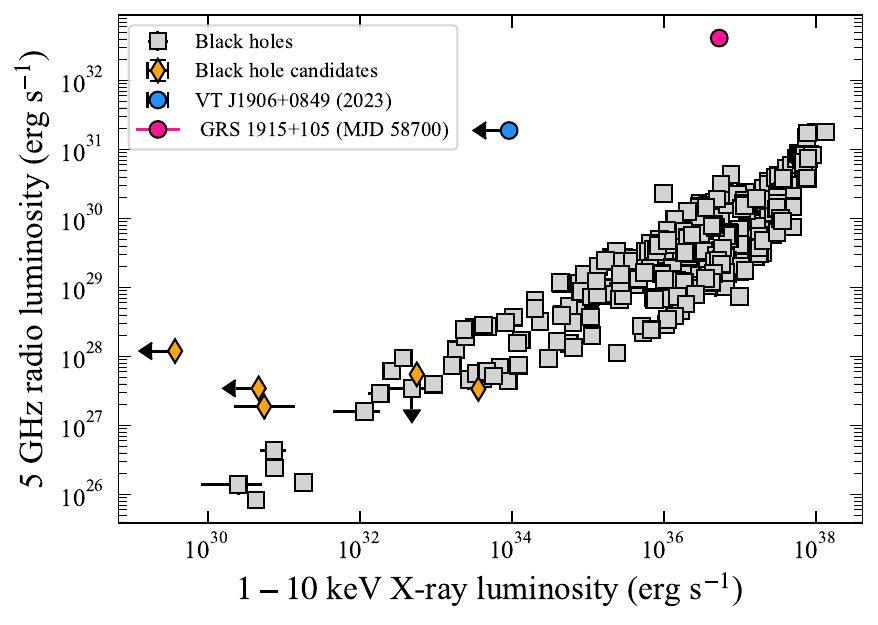}
    \caption{Radio/X-ray correlation for hard state candidate (white) and confirmed (gray) black hole X-ray binaries from \cite{arash_bahramian_2022_7059313}. The pink circle represents GRS 1915+105 during a brief radio-bright, X-ray obscured state \citep{2021MNRAS.503..152M}. VT J1906+0849 is shown as a blue circle. }
    \label{fig:lrlx}
\end{figure}

Despite these similarities, an X-ray binary interpretation faces several important challenges. Most notably, no X-ray counterpart is detected. The Swift-XRT observation implies a $3\sigma$ upper limit on the unabsorbed X-ray luminosity of $L_X\leq9\times10^{33}$ erg s$^{-1}$. Using a 5 GHz radio flux density measured within three months of the X-ray non-detection, VT J1906+0849 lies well above the canonical hard state radio/X-ray correlation  (Figure \ref{fig:lrlx}) established for accreting black hole X-ray binaries \citep[e.g.,][]{2000A&A...359..251C,2003A&A...400.1007C,2003MNRAS.344...60G,2012MNRAS.423..590G,2014MNRAS.445..290G}. Although the radio and X-ray measurements are separated by a few months, VT J1906+0849 exhibits only modest radio variability on comparable month-long timescales, far smaller than the offset from the canonical radio/X-ray correlation. The non-simultaneity of the observations is therefore unlikely to explain the discrepancy.

Radio-bright, X-ray suppressed states are known in individual X-ray binaries. In particular, GRS 1915+105 entered an extended obscured phase in which strong radio activity persisted while the directly observed X-ray emission remained faint, plausibly due to local absorption by a dense disk wind or geometrically thick accretion flow \citep{2021MNRAS.503..152M,2023A&A...680L..16S}. Shorter-lived examples of strong local obscuration and wind-dominated evolution have also been observed during the outburst of V404 Cyg \citep{2016Natur.534...75M,2017MNRAS.468..981M}. These systems demonstrate that local obscuration can temporarily decouple the observed radio and X-ray luminosities. However, applying this explanation to VT J1906+0849 would require an obscured, radiatively inefficient, or otherwise X-ray-suppressed accretion configuration to remain viable throughout a radio transient that has evolved over $\sim20$ yr. An alternative possibility is that the observed radio luminosity does not trace the instantaneous compact jet power, but instead arises from long-lived ejecta or an outflow-environment interaction.

\paragraph{Jet-wind interaction}
One possible explanation for the compact radio emission from VT J1906+0849 is that a continuously powered synchrotron outflow is confined through interaction with a dense disk wind. A useful comparison is SS 433, a persistent supercritical accretor in which a geometrically thick accretion disk launches a massive wind, produces a funnel-like geometry and powers narrow baryonic jets \citep{1979Natur.279..701A,1984ARA&A..22..507M,2002PASJ...54..253O,2004ASPRv..12....1F}. This is in contrast to all other X-ray binaries, where the disk wind is present in the jet-free soft state \citep{2012MNRAS.422L..11P}, and may be a signature of a supercritical accretor.

The observed properties of SS 433 depend strongly on viewing angle: the inner accretion flow is substantially obscured when viewed away from the funnel axis, while reprocessed and scattered emission emerges from the wind and funnel walls. We consider whether a related but much younger analog of SS 433's jet-wind interaction could explain VT J1906+0849's compact morphology and X-ray faintness.

Several observations suggest that the radio-emitting plasma is not freely expanding. Under the synchrotron self-absorption interpretation, the equipartition radius remains $\sim10^{14}$ cm over several years (Table \ref{tab:energetics}), corresponding to a characteristic expansion velocity of $\lesssim10$ km s$^{-1}$. The VLBI measurements provide an independent constraint. Between 2010\footnote{For the 2010 epoch, we use the \cite{2025ApJS..276...38P} calibrated measurement because the longest baselines were excluded from our re-analysis.} and 2025, the deconvolved major axis size at 8.4 GHz does not increase significantly. The minor axis changes by only $\approx0.4$ mas, corresponding to a radial expansion of order 3 AU and $v_{\rm exp}\lesssim1$ km s$^{-1}$ at the fiducial distance of 15 kpc. Scatter broadening may contribute to the measured VLBI size, so these values should be interpreted cautiously. Nevertheless, both the equipartition and VLBI estimates indicate that the radio-emitting region remains compact over many years. The synchrotron cooling timescale of several years further suggests that the radiating particles must be continuously replenished or reaccelerated. 

The broad near-infrared line may provide evidence for the confining medium. Its centroid remains stable across multiple epochs, while its width and luminosity vary. If the line is associated with blueshifted Br$\gamma$, its centroid implies a bulk line of sight velocity of $\approx-2600$ km s$^{-1}$, which is comparable to its measured FWHM of $\approx1400-2220$ km s$^{-1}$. A stable blueshifted centroid would favor a persistent asymmetric outflow geometry in which the approaching side of a fast wind is preferentially visible, while the receding side is obscured or substantially fainter. We therefore adopt a representative wind velocity of $v_w\approx2000$ km s$^{-1}$, motivated jointly by the candidate Br$\gamma$ blueshift and the observed line width, although the precise interpretation depends on the line identification and geometry.

The effective internal pressure of the synchrotron-emitting region is 
$$P_{\rm syn}\approx \frac{u_{\rm int}}{3}\approx \frac{E_{\rm tot}^{(\rm min)}}{V}$$ where $u_{\rm int}$ is the internal energy density in synchrotron-emitting particles and the magnetic field, $V=4\pi R^3\phi/3$ is the volume of the emitting region, and $\phi\sim1$ is the volume filling fraction. We find $P_{\rm syn}\sim10^{-2}$ dyn cm$^{-2}$ in 2019. If the radio-emitting plasma is confined by the ram pressure of a disk wind, 
$$P_{\rm ram}\approx\rho v_w^2\approx \mu nm_pv_w^2$$ 
where $\rho$ is the wind mass density, $n$ is the wind number density, and $\mu$ is the mean molecular weight per particle, then the wind density must be high enough that $P_{\rm ram}>P_{\rm syn}$, or $n>P_{\rm syn}/(\mu m_p v_w^2)$. Assuming $\mu\sim1$, we estimate a required density of $\gtrsim7\times10^5$ cm$^{-3}$. 

If confinement occurs near the equipartition radius, the required mass-loss rate is 
$$\dot{M}_w\approx 4\pi R^2 v_w\mu nm_p $$
or a few $\times10^{-7} M_\odot$ yr$^{-1}$, which is comparable to the black hole accretion rate estimated from the radio luminosity-jet power scaling. The inferred mass-loss rate is substantially lower than the canonical value of $\sim10^{-4} M_\odot$ yr$^{-1}$ inferred for SS 433 \citep{1981VA.....25...95V,1997Ap&SS.252..439F,2002ASPC..271..369F}. However, it is still sufficiently high to motivate a less extreme analog in which a dense, structured disk wind confines or shocks the synchrotron-emitting plasma.

The X-ray non-detection remains an important challenge for this interpretation. A Compton-thick column requires a hydrogen column density of $N_H\geq1.5\times10^{24}$ cm$^{-2}$ \citep{2004ASSL..308..245C}. If the AU-scale radio-confining gas had a uniform density of $\approx7\times10^5$ cm$^{-3}$, the required path length would be $\approx2\times10^{18}$ cm, or 0.7 pc. This is much larger than the radio-emitting region. The AU-scale gas responsible for confining the radio source is therefore unlikely to obscure the central engine by itself.

This does not rule out wind obscuration. A steady outflow is not expected to have a uniform density; for an approximately radial wind, mass conservation gives $n(r)\propto\dot{M}_w/[r^2v_w(r)]$, which reduces to $n(r)\propto r^{-2}$ outside the accelerating region \citep{1975MNRAS.170...41W}. The wind density increases sharply toward the central engine, and the integrated absorbing column is dominated by material at the smallest radii, which may be $\gg10^5$ cm$^{-3}$. The radio and X-ray emission may also traverse different columns. The radio emission could escape from a larger jet-wind interaction region, while the compact X-ray source remains hidden by a denser inner wind or equatorial outflow, analogous to the wind-obscured geometry inferred for SS 433 \citep{2004ASPRv..12....1F,2021MNRAS.506.1045M}.

In this picture, VT J1906+0849 contains an accreting compact object surrounded by a dense and asymmetric disk wind. A compact synchrotron outflow in the form of a jet interacts with this wind on AU scales, producing a slowly evolving radio source that must be continuously powered. The approaching side of the wind may produce the persistent blueshifted near-infrared line, while a denser inner component suppresses direct X-ray emission and the redshifted component of the disk wind. This framework naturally links the compact radio morphology, the evidence for ongoing particle replenishment, the broad emission line, and the X-ray faintness, while allowing VT J1906+0849 to be substantially less extreme than SS 433.

Several open questions remain. The identification of the near-infrared line as blueshifted Br$\gamma$ is not secure. The absence of additional hydrogen and helium lines is also notable, particularly given the rich emission and absorption line spectrum of SS 433 \citep{2002ApJ...566.1069G,2002ASPC..271..369F,2004ASPRv..12....1F,2017ApJ...841...79R}. The absence of, e.g., moving lines associated with the baryonic jets in SS 433 is expected in VT J1906+0849 if the jets are leptonic. Multi-epoch, high-resolution ($R\gtrsim10,000$) near-infrared spectroscopy would help determine whether the broad line traces a stable wind geometry, orbital motion, or a more complex outflow. A deep hard X-ray observation would test whether an absorbed or reflected central source emerges above the soft X-ray band, and is the natural next step in the study of VT J1906+0849.

\section{Summary}

We have presented a multiwavelength study of VT J1906+0849, the brightest radio transient discovered in VLASS to date. The source is a distant Galactic synchrotron transient that has remained radio-bright for approximately two decades while exhibiting an evolving spectral turnover frequency and little apparent expansion. Together with the persistent broad near-infrared emission line, these properties favor sustained accretion onto a compact object rather than an impulsive explosion or passively evolving nebula. We propose that the radio emission is powered by the interaction of a continuously replenished synchrotron outflow with a dense, asymmetric disk wind, which may also contribute to the X-ray non-detection. VT J1906+0849 highlights the potential for wide-field radio surveys to uncover a population of obscured Galactic accretors that are difficult to identify at other wavelengths.

\begin{acknowledgments}

The authors thank Mark Reid, Gregory R. Sivakoff, and James Miller-Jones for helpful discussions. We are also grateful to Lynne A. Hillenbrand for obtaining the 2026 Keck-II/NIRES spectrum. Funded in part by Schmidt Sciences and the United States Israel Binational Science Foundation (Grant number 2020203). A.H. is grateful for the support by the Israel Science Foundation (ISF grant 1679/23) and by the United States-Israel Binational Science Foundation (BSF grant 2020203).

\end{acknowledgments}

\software{\texttt{Astropy} \citep{astropy_2013,astropy_2018,astropy_2022}, \texttt{NumPy} \citep{harris2020array}}


\appendix

\section{Proper motion}\label{sec:proper_motion}

Proper motion measurements from high-precision radio interferometry can provide indirect constraints on the distance to compact sources by comparing observed transverse motions to expectations from Galactic rotation. In the absence of significant intrinsic motion (e.g., from natal kicks), a low observed proper motion implies a large distance, since the angular motion resulting from Galactic rotation scales inversely with distance. This technique was applied to infer minimum distances for X-ray binaries such as Cygnus X-3 and GRS 1915+105 \citep{2023ApJ...959...85R}, where radio VLBI observations placed tight constraints on proper motion, thereby excluding nearby solutions and favoring kinematic distances of several kiloparsecs. We adopt a similar approach to constrain the location of VT J1906+0849.

To interpret the observed proper motion constraint in the context of Galactic dynamics, we modeled the expected apparent motion of a source moving in a circular orbit around the Galactic center as a function of distance. We assume the source follows pure Galactic rotation and computed its expected transverse motion relative to the Sun at each line of sight from 2 to 30 kpc. We use the \texttt{MWPotential2014} Galactic potential from the \texttt{galpy} package \citep{2015ApJS..216...29B} to determine the local circular velocity at each Galactocentric radius, adopting the Galactocentric frame parameters of $R_0=8.34$ kpc and $V_0=240$ km s$^{-1}$ \citep{2014ApJ...783..130R}. The resulting space velocity of the object, transformed into the observer's frame and projected onto Galactic coordinates, yields the predicted proper motion components $\mu_l\cos b$ and $\mu_b$. These can then be compared to observations to constrain the distance to VT J1906+0849.

\begin{figure} 
    \centering
    \includegraphics[width=0.5\textwidth]{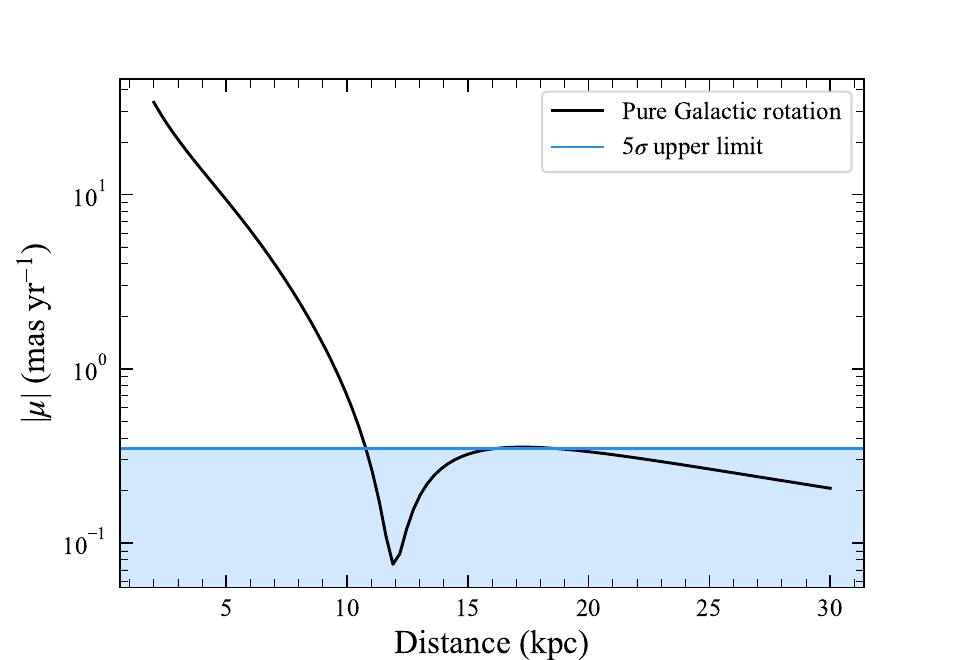}
    \caption{Expected total proper motion $|\mu|$ towards VT J1906+0849 from pure Galactic rotation as a function of distance (black curve). The $5\sigma$ upper limit on measured proper motion between the 2010 and 2025 VLBA X-band epochs is shown by the horizontal blue line, with the shaded region indicating values consistent with the observation.}
    \label{fig:proper_motion}
\end{figure}

Between the 2010 and 2025 X-band VLBA epochs, VT J1906+0849 shows only marginal evidence for proper motion: we measure $-0.017\pm0.009$ mas yr$^{-1}$ in right ascension, while the declination shift is consistent with zero, $-0.07\pm0.08$ mas yr$^{-1}$, and is treated as an upper limit. The resulting total proper motion is $|\mu|=0.07\pm0.07$ mas yr$^{-1}$, for which we adopt a conservative $5\sigma$ upper limit of $|\mu|\lesssim0.35$ mas yr$^{-1}$. Figure \ref{fig:proper_motion} compares this measurement to the expectation from Galactic rotation at the source's position ($l\approx42.5^\circ,b\approx0.74^\circ$). Assuming negligible peculiar motion, this upper limit is consistent with a Galactic source at $d\gtrsim11$ kpc, or an extragalactic source.

We note this distance constraint assumes VT J1906+0849 follows pure Galactic rotation with negligible peculiar proper motion. In practice, disk objects may exhibit additional radial and azimuthal peculiar velocities that can cancel or enhance the apparent transverse motion. 

\section{Equipartition properties}\label{sec:ssa_equipartition}

If $p$ is the electron energy index where $N(E)dE=N_0 E^{-p}dE$, $F_{\nu_{\rm pk}}$ is the flux density at the turnover frequency $\nu_{\rm pk}$, $k$ is the  ratio of heavy-particle energy
to relativistic electron energy, $\varphi$ is the volume filling factor, and $d$ is distance, then the equipartition radius, $R$, and magnetic field, $B$, are 
$$R=\left[\frac{(2p+2)C_e}{11C_B}\right]^{1/(2p+13)},\ B=AR^4$$ 
where
$$C_e\equiv(1+k)\frac{4\pi}{3}\varphi\left(\frac{\nu_{\rm pk}}{2c_1}\right)^{(p+4)/2}\frac{\int E^{1-p}dE}{c_6(p)A^{(p+2)/2}},\ C_B\equiv\frac{\varphi}{6}A^2,\ A\equiv\left[\frac{\pi R^2}{F_{\nu_{\rm pk}}d^2}\frac{c_5(p)}{c_6(p)}\left(\frac{\nu_{\rm pk}}{2c_1}\right)^{5/2}(1-e^{-1})\right]^2.$$ 
The minimum total energy is 
$$ E_{\rm total}^{(\min)}=(1+k)\left(1+\frac{2p+2}{11}\right)E_e=(1+k)\left(1+\frac{2p+2}{11}\right)E_e=\left(1+\frac{11}{2p+2}\right)E_B $$
where $E_e=\frac{4\pi}{3}\varphi R^3\int E^{1-p}dE$ and $E_B=(\varphi R^3B^2)/6.$ The synchrotron timescale is $\tau=c_{12}(p)B_\perp^{-3/2}$. The constants $c_1$, $c_5(p)$, $c_6(p)$, and $c_{12}(p)$ are provided by \cite{1970ranp.book.....P}. We assume a plasma with a ratio of electrons to protons of $k\sim1$ emits at frequencies $10^7$ to $10^{11}$ Hz in a spherical volume with a volume filling factor $\varphi\sim1$.



\bibliography{biblio}{}
\bibliographystyle{aasjournalv7}



\end{document}